\documentclass[11pt,twocolumn,aps,prl,reprint]{revtex4-1}

\usepackage{braket}
    
\usepackage{graphicx,amsmath,amssymb,braket}

\begin{document}

\title{Plasmons in tunnel-coupled graphene layers: backward waves with quantum cascade gain}
\author{D. Svintsov$^{1}$, Zh. Devizorova$^{2,3}$, T. Otsuji$^{4}$, and V. Ryzhii$^{4}$}
\affiliation{$^{1}$Laboratory of Nanooptics and Plasmonics, Moscow Institute of Physics and Technology, Dolgoprudny 141700, Russia}
\affiliation{$^{2}$Department of Physical and Quantum Electronics, Moscow Institute of Physics and Technology, Dolgoprudny 141700, Russia}
\affiliation{$^{3}$Kotelnikov Institute of Radio Engineering and Electronics, Russian Academy of Science, Moscow, 125009 Russia}
\affiliation{$^{4}$Research Institute of Electrical Communication, Tohoku University,  Sendai 980-8577, Japan}
 
\begin{abstract}
Plasmons in van der Waals heterostructures comprised of graphene and related layered materials demonstrate deep subwavelength confinement and large propagation length. In this letter, we show that graphene-insulator-graphene tunnel structures can serve as plasmonic gain media. The gain stems from the stimulated electron tunneling accompanied by the emission of coherent plasmons under interlayer population inversion. The probability of tunneling with plasmon emission appears to be resonantly large at certain values of frequency and interlayer voltage corresponding to the transitions between electron states with collinear momenta -- a feature unique to the linear band structure of graphene. The dispersion of plasmons undergoes a considerable transformation due to the tunneling as well, demonstrating negative group velocity in several frequency ranges.

\end{abstract}

\maketitle

The ultrarelativistic nature of electrons in graphene gives rise to the uncommon properties of their collective excitations -- surface plasmons (SPs)~\cite{Graphene_plasmonics-1,Graphene_plasmonics-2,Das_Sarma_Plasmons}. The deep subwavelength confinement~\cite{Graphene_plasmonics-2}, the unconventional density dependence of frequency~\cite{Ryzhii-plasmons,Das_Sarma_Plasmons}, and the absence of Landau damping~\cite{Ryzhii-plasmons} are probably their most well-known features.  Among more sophisticated predictions there stand the existence of transverse electric plasmon modes~\cite{Mikhailov_new_mode} and quasi-neutral electron-hole sound at the charge neutrality~\cite{Our-hydrodynamic, New_plasmon_mode}. It was not until the discovery of van der Waals heterostructures that the truly low-loss SPs supported by graphene with propagation length to wavelength ratio reaching 25 could be observed~\cite{Koppens_nano_imaging_hBN}. The respective SP damping rate is order of $0.5$ ps, and it can be potentially compensated by the plasma instabilities~\cite{Polini_FET, Our_NLHD} or stimulated plasmon emission in pumped samples~\cite{Dubinov_JPCM,Rana_IEEE}.

In this letter, we demonstrate theoretically that the resonant tunneling structures comprised of parallel graphene layers can act as plasmonic gain media by themselves. Apparently, the negative differential resistance (NDR) in tunnel diodes can give rise to the self-oscillation in electrical circuits, but the extension of this concept to the self-excitation of plasmons is not trivial~\cite{Feiginov_Volkov,Ryzhii_Shur_JJAP,Berardi_APL}. In addition, the weak NDR observed in the {\it static} current-voltage curves of graphene tunnel diodes is insufficient to replenish the plasmon losses, which calls for the stability of electron plasma~\cite{Dynamic_effects}. However, the dynamic and non-local effects in the tunnel conductivity can radically change the picture of plasmon propagation.

\begin{figure}[ht]
\includegraphics[width=0.8\linewidth]{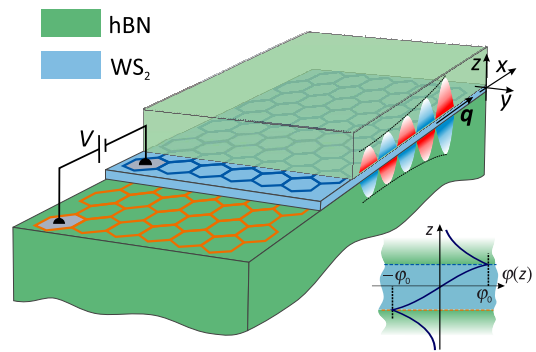}
\caption{\label{Structure} Schematic view of the double graphene layer encapsulated in hexagonal boron nitride (hBN) overlaid by the image of acoustic SP amplified by the tunneling. Inset: spatial distribution of electric potential $\varphi(z)$ in acoustic SP mode.
}
\label{F1}
\end{figure}

 We calculate the dynamic tunnel conductivity of double graphene layer biased by voltage $V$ and show that its real part is negative at frequencies $\omega < e V/\hbar$, which is a consequence of the interlayer population inversion. More surprisingly, the negative tunnel conductivity possesses sharp resonances at certain frequencies and wave vectors $q$, despite the absence of any quantum-confined subbands in the structure. Instead, the resonances emerge due to the prolonged tunneling interaction between the states with collinear momenta in neighboring graphene layers. The singularities in the tunnel  conductivity emerge at a series of lines on the $\omega-q$ plane, whose pattern is especially rich in the presence of interlayer twist. At finite bias $V$, the dispersion of acoustic SPs  does not develop a low-frequency tunnel gap, as opposed to SPs in coupled layers of massive electrons in equilibrium~\cite{DasSarma-PRL-tunnel-plasmon}. Instead, the SP spectrum develops an anticrossing with the tunnel resonances and demonstrates the parts with negative group velocity. At the same time, the dispersion passes quite close to the tunnel resonances, and the tunneling gain can exceed the plasmon loss due to both inter- and intraband SP absorption.
 
We start with the generalization of the acoustic plasmon dispersion law~\cite{Hwang_PRB_2GL,Voltage_controlled} accounting for the tunneling~\cite{DasSarma-PRL-tunnel-plasmon} between parallel layers of electrically doped graphene shown in Fig.~\ref{F1}. For equal electron and hole doping of opposite layers, the dispersion equation reads (see Supporting information, Sec.~I)
\begin{equation}
\label{Acoustic_mode}
1+\frac{2\pi i q}{\omega \kappa}\left[\sigma_\parallel ({\bf q},\omega) + \frac{2G_\bot ({\bf q},\omega )}{q^2}\right]\left( 1-e^{-qd} \right)=0,
\end{equation}
where $d$ and $\kappa$ are the thickness and permittivity of interlayer dielectric,  $\sigma_\parallel$ is the in-plane graphene conductivity, and $G_\bot$ is the tunnel conductivity, the proportionality coefficient between the tunnel current density and the interlayer voltage drop [units: Ohm$^{-1}$m$^{-2}$].

An only missing ingredient required for the analysis of surface plasmon modes is the expression for the high-frequency non-local tunnel conductivity $G_\bot (\bf{q},\omega )$. The theoretical studies of the latter have been limited to the DC~\cite{Brey_PRA,Vasko_PRB} or local ($q=0$) AC cases~\cite{Ryzhii_DGL_laser}. Here, we consider the linear response of voltage-biased graphene layers to the propagating acoustic plasmon whose electric potential $\delta\varphi(z)e^{iqx-i\omega t}$ is highly nonuniform (see inset in Fig.~\ref{F1}). The electrons in tunnel-coupled graphene layers are described with the tight-binding Hamiltonian
\begin{equation}
\label{Hamiltonian}
\hat{H}_0=\left( \begin{matrix}
   {{{\hat{H}}}_{G+}} & {\hat{\mathcal{T}}}  \\
   {{{\hat{\mathcal{T}}}}^{*}} & {{{\hat{H}}}_{G-}}  \\
\end{matrix} \right),
\end{equation}
where the blocks ${\hat H}_{G\pm} = v_0 {\boldsymbol \sigma} {\hat {\bf p}} \pm {\hat I}\Delta/2$ stand for isolated graphene layers, $v_0 = 10^6$ m/s is the Fermi velocity, $\Delta$ is the voltage-induced energy spacing between the Dirac points, $\hat{\bf p}$ is the in-plane momentum operator, $\hat I$ is the identity matrix, and $\hat{\mathcal T} = \Omega {\hat{I}}$ is the tunneling matrix. Such model of tunnel couping applies to the AA-aligned graphene layers~\cite{Brey_PRA,Twisted_GBL}; the effects of layer twist will be briefly addressed in the end of paper.

\begin{figure}[t]
\includegraphics[width=0.85\linewidth]{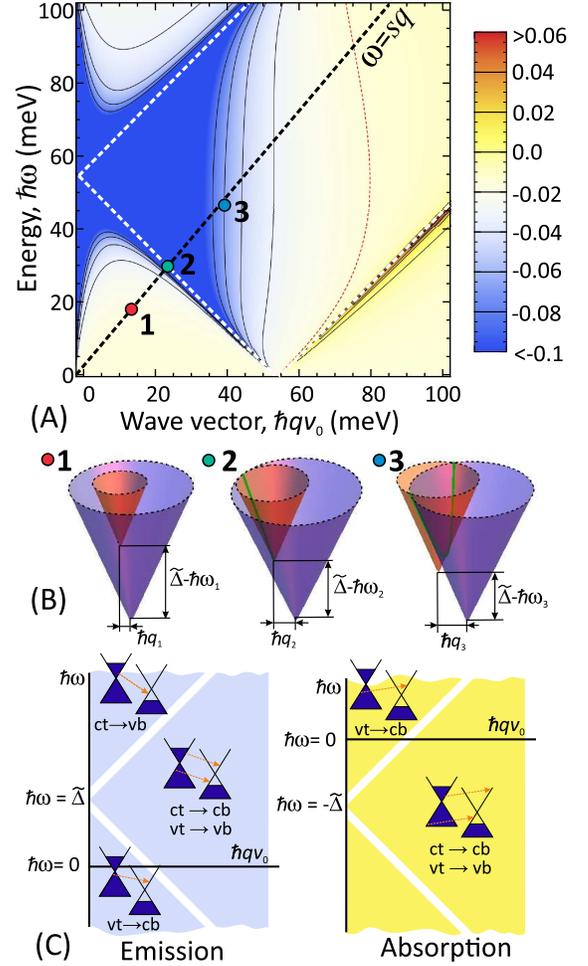}
\caption{\label{F2} 
(A) Color map of the tunnel conductivity, $2 {\rm Re}G_{\bot}/q^2$ (units of $e^2/\hbar$), calculated at temperature $T=77$ K and interlayer voltage $V=0.2$ V. Dielectric layer is 3 nm WS$_2$ (effective mass $m^* = 0.28 m_0$, conduction band offset to graphene $U_b = 0.4$ eV~\cite{Band_parameters_MOS2}). Red dashed line corresponds to the zero conductivity, black dashed line shows the dispersion of acoustic SP in the absence of tunneling (B) Band diagrams illustrating available electron states for plasmon-assisted tunneling at different frequencies and wave vectors. Position (2) corresponds to the resonant collinear tunneling. (C) Map of the frequency- and wave vector ranges, for which the interlayer transitions accompanied by the $(\omega,q)$-quantum emission and absorption are possible. }
\end{figure}

The evaluation of interlayer conductivity is based on the solution of the quantum Liouville  equation for the electron density matrix followed by the evaluation of the expectation value of the current operator. The calculations are conveniently performed in the basis of $\hat H_0$ - eigenstates labelled by the in-plane momentum ${\bf p}$, the index $s=\{c,v\}$ for the conduction and valence bands, respectively, and the number $l=\pm 1$ governing the $z$-localization of the wave function (Supporting information, Sec.~II). At strong bias, $\Delta \gg \Omega$, the state with $l=+1$ ($-1$) is localized primarily on the top (bottom) layer. Upon lowering the bias, the state with $l=+1$ ($-1$) becomes odd (even) with respect to $x$. The states' energies are $\varepsilon^{ls}_{\bf p} = s v_0 p + l \tilde\Delta/2$, where $\tilde\Delta = \sqrt{4\Omega^2 + \Delta^2}$ is the spacing between levels in the voltage-biased tunnel-coupled wells~\cite{vasko_book}. 

The outlined scheme leads us to the following expression for the components of conductivity (Supporting information, Sec.~III):
\begin{multline}
\label{In-plane}
{\sigma_{\parallel }}( {\bf q},\omega )=-ig\frac{e^2}{\hbar }{S_{++}}\cos\theta_M \times\\
\sum\limits_{ss'\bf{p}}{\frac{{{| {\bf v}_{{\bf p}{\bf p}'}^{ss'} |}^{2}}}{\varepsilon _{{\bf p}_-}^{s’}-\varepsilon _{{{\bf p}_+}}^s}\frac{f_{{\bf p}_+}^s-f_{{\bf p}_-}^{s'}}{\varepsilon _{{\bf p}_+}^s-\varepsilon _{{\bf p}_-}^{s’}-\left( \hbar\omega +i\delta  \right)}},
\end{multline}

\begin{multline}
\label{Out-plane}
G_\bot ( {\bf q},\omega )=-ig\frac{e^2}{2\hbar }{S_{\pm}}\sin\theta_M\times\\
\sum\limits_{\begin{smallmatrix} 
 l\ne l' \\ 
 ss'\bf{p} 
\end{smallmatrix}}{{{| u_{{\bf p}{\bf p}'}^{ss'} |}^{2}}\frac{\varepsilon _{{\bf p}_-}^{s'l'}-\varepsilon_{{\bf p}_+}^{sl}}{\varepsilon_{{\bf p}_+}^{sl}-\varepsilon _{{\bf p}_-}^{s'l'}-\left( \hbar\omega +i\delta  \right)}}\left( f_{{\bf p}_+}^{sl}-f_{{{\bf p}_-}}^{s'l'} \right).
\end{multline}
%\begin{equation}
%\label{Mixing}
%\sigma_{q\omega}+\frac{G_{q\omega}}{q^2}={S_{++}}\cos {\theta_{M}}{\sigma_\parallel}( {\bf q},\omega  ) + {S_\pm}\sin {\theta_{M}}{\sigma_\bot} ( \bf{q},\omega ),
%\end{equation}
Above, $g=4$ is the spin-valley degeneracy factor, $\theta_M$ is the 'mixing angle' characterizing the strength of the tunnel coupling, $\sin \theta_M = 2\Omega/\tilde\Delta$; $S_{++}$ and $S_{\pm}$ are the overlap integrals of plasmon potential (normalized by its on-plane value) and the $z$-components of $H_0$ eigenfunctions~\footnote{Note that the in-plane conductivity is also renormalized due to the delocalization of electron wave function outside of graphene and nonuniformity of plasmon field, $S_{++}<1$}; ${\bf p}_{\pm} \equiv {\bf p}\pm \hbar {\bf q}/2$, $u_{{\bf p}{\bf p}'}^{ss'}$ and ${\bf v}_{{\bf p}{\bf p}'}^{ss'}$ are the matrix elements of projection and velocity operators between chiral states $\ket{{\bf p}s}$ and $\ket{{\bf p}'s'}$ in a single graphene layer. Finally, $f^{sl}_{\bf p}$  and $f^{s'l'}_{{\bf p}'}$ are the occupation functions of the respective states, which are assumed to be the Fermi functions shifted by $eV$ in the energy scale for the opposite $l$-indices. 

%The case of weak bias will be of no interest for us further -- we just note that if the layers are not doped intentionally, the plasmons in this case will be strongly damped due to the interband transitions. 

The first peculiarity of Eq.~(\ref{Out-plane}) worth discussing is the negative value of the real part of tunnel conductivity at frequencies $\omega < e V/\hbar$. This negativity implies that the interlayer transitions accompanied by the emission of the field quantum $(\omega,{\bf q})$ are more probable than the inverse absorptive transitions. The band filling providing the negative tunnel conductivity can be viewed as an interlayer population inversion similar to that in the quantum cascade lasers~\cite{Kazarinov_Suris,Capasso_Science}. The frequency- and wave vector dependence of $2 {\rm Re}G_\bot /q^2$ is shown in Fig.~\ref{F2}A, where the 'cold' colors stand for the negative and 'warm' colors for the positive conductivity. An analysis of energy-momentum conservation reveals distinct regions on the frequency-wave vector plane, where different types of tunnel transitions with emission or absorption of the field quantum are relevant, see Fig.~\ref{F2}C. Among those, the most pronounced is the interlayer intraband emission allowed within the quadrant $qv_0 \ge |\tilde{\Delta}/\hbar - \omega|$. The interband transitions are generally weaker due to the small overlap of chiral wave functions of different bands~\cite{Chiral_tunneling} (see Supporting information, Sec.~IV for analytical approximations to the tunnel conductivity).

A distinct feature of the tunnel conductivity readily observed in Fig.~\ref{F2}A is its large absolute value near a series of lines $q v_0 = |\omega \pm \tilde \Delta/\hbar|$. The origin of these resonances can be explained by analyzing the possible electron states involved in plasmon-assisted tunneling at different frequencies and wave vectors, Fig.~\ref{F2}B. To be precise, we focus on the interlayer intraband tunneling. Above the resonance, at $qv_0 > \omega - \tilde\Delta/\hbar$, the electrons capable of tunneling occupy a hyperbolic cut of the mass shell in graphene (case B3 in Fig.~\ref{F2}). With decreasing the frequency and wave vector, the hyperbola degenerates into a straight line (case B2). At this point, the tunneling occurs between states with collinear momenta and equal velocities -- hence, their interaction would last for an infinitely long time were there no carrier scattering~\footnote{Alternatively, the singularities in the tunnel conductivity can be traced back to the van Hove singularities in the joint density of states~\cite{Brey_PRA}}. At even lower frequencies (case B1), the intraband transitions are impossible, but the weaker interband tunneling sets in. In the absence of scattering, the collinear tunneling singularities are square-root, 
\begin{equation}
{\rm Re}G^{\rm intra}_\bot \propto [q^2v_0^2 - (\tilde\Delta/\hbar - \omega)^2]^{-1/2},   
\end{equation}
similar to the absorption singularities at the onset of the Landau damping
\begin{equation}
 {\rm Re}\sigma^{\rm intra}_\parallel \propto [q^2v_0^2 - \omega^2]^{-1/2}.   
\end{equation}
The actual value of the resonant conductivity is limited by electron-acoustic phonon scattering at low electron energies~\cite{Principi_plasmon_loss_hBN}. We account for it by replacing the delta-peaked spectral functions of individual electrons in Eqs.~(\ref{In-plane}) and (\ref{Out-plane}) with Lorentz functions of proper width~\cite{Brey_PRA}. With the scattering rate $\tau_{\rm tr}^{-1} \simeq (2 \div 8) \times 10^{-11}$~s$^{-1}$ at $T= 77\div300$ K~\cite{Vasko-Ryzhii,PRL_Bolotin} and electron density $n=5 \times 10^{11}$ cm$^{-2}$, the tunnel resonances remain pronounced even at room temperature.

\begin{figure}[t]
\includegraphics[width=0.9\linewidth]{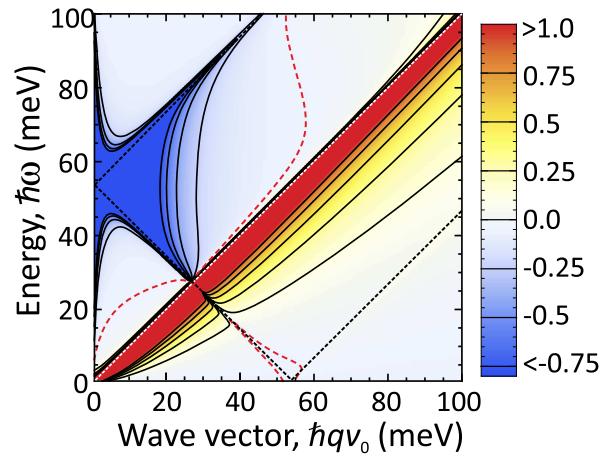}
\caption{\label{F3} 
Space-time dispersion of the effective conductivity ${\rm Re}[\sigma_\parallel + 2G_\bot/q^2]$ governing the damping (or gain) of acoustic SPs. The values are normalized by $e^2/\hbar$. The structure parameters are the same as in Fig.~\ref{F2}. The contour of zero conductivity is highlighted with red dashed line
}
\end{figure}

Were there no collinear singularities in the electron tunneling, its effect on plasmon spectra and damping would be small. The presence of resonances suggests the possibility of the net plasmon gain and strong renormalization of plasmon dispersion. The attenuated or amplified character of SP propagation is governed by the sign of the 'effective conductivity' ${\rm Re}[\sigma_\parallel + 2G_\bot/q^2]$ plotted in Fig.~\ref{F3}. The proximity of acoustic SP velocity to the Fermi velocity at small $d$~\cite{Ryzhii-plasmons,Polini-soundarons} antagonizes the net gain as both inter- and intraband absorption are large near $\omega = q v_0$~\cite{Das_Sarma_Plasmons} (at $d=2.5$ nm, the velocity of SPs not perturbed by the tunneling is $s\approx 1.1 v_0$). On the other hand, a large ratio of transverse and in-plane electric fields in the acoustic mode, ${\cal E}_\bot/{\cal E}_\parallel = 2(qd)^{-1} \gg 1$ facilitates the tunneling gain compared to the in-plane absorption. The competition of these factors results in the emergence of relatively broad frequency-wave vector ranges (encircled by red lines in Fig.~\ref{F3}) where the real part of the effective conductivity is negative and the net SP gain is possible.

\begin{figure}[ht]
\includegraphics[width=0.9\linewidth]{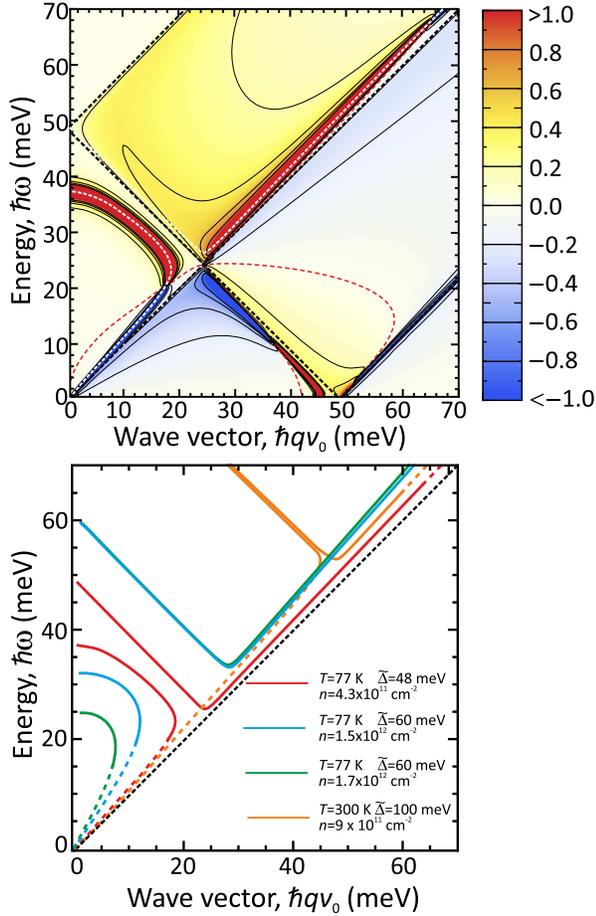}
\caption{\label{F4} 
(A) Spectral function of acoustic SPs calculated for $2.5$ nm WS$_2$ barrier layer, $T = 77$ K and $V=0.2$ V. The plasmon spectrum develops an anticrossing with the collinear tunneling resonance (B) Plasmon spectra calculated for different temperatures, electron densities and level spacing of tunnel coupled layers $\tilde\Delta$. Dashed parts of the spectra correspond to the damped and solid parts -- to the amplified plasmons. Black dashed line is $\hbar\omega = q v_0$
}
\end{figure}

The square-root singularities in ${\rm Re}G_\bot^{\rm intra}$ above the threshold of interlayer interband transitions are mirrored into the singularities in ${\rm Im}G_\bot^{\rm intra}$ below the threshold, which is proved by the virtue of Kramers-Kronig relations. A similar situation holds for the in-plane conductivity, whose imaginary part is positive and resonantly large above the domain of Landau damping, i.e. at $\omega \rightarrow qv_0 + 0 ^+$. The interplay of two singular conductivities (in-plane and out-of-plane) results in the 'locking' of the long-wavelength part of plasmon spectrum in the domain $(\omega \ge q v_0)\cup (\omega \le \tilde\Delta/\hbar - qv_0)$. This is clearly seen in the plot of acoustic plasmon spectral function ${\cal S}({\bf q},\omega)$, the imaginary part of the inverse of Eq.~(\ref{Acoustic_mode}), Fig.~\ref{F4}A. With increasing the frequency, the plasmon peak develops an anticrossing with the resonance in the tunnel conductivity. The group velocity of acoustic SPs in the vicinity of tunnel resonance is negative and close to $-v_0$. Above the resonance, the initially linear SP dispersion remains almost unperturbed, though the excitations are still amplified but not attenuated.

\begin{figure}[ht]
\includegraphics[width=0.85\linewidth]{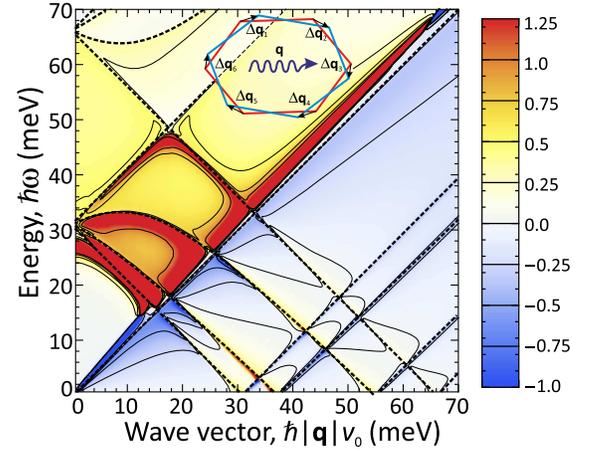}
\caption{\label{F5} 
Spectral function calculated for acoustic SPs in twisted layers (twist angle $\theta_T = 0.57^\circ$) propagating along the misalignment vector in one pair of valleys. Inset shows the positions of K-points in the reciprocal space for twisted graphene layers
}
\end{figure}

The effects of tunneling on SP dispersion are generally more pronounced at low levels' spacing $\tilde\Delta$ and high carrier density $n$. These quantities can be addressed independently in gated double layers. By fixing $\tilde \Delta$ and increasing the carrier density, one can achieve a large enhancement of the SP velocity below the tunnel resonance, as shown in Fig.~\ref{F4}B, and at some critical density the long-wavelength branch of SP dispersion can disappear at all. Such behaviour contrasts to the plasmons in coupled layers of massive electrons at equilibrium, where the large negative value of ${\rm Im}G_\bot$ at $\hbar\omega < 2 \Omega$ result in a gapped SP dispersion~\cite{DasSarma-PRL-tunnel-plasmon,Tunneling_plamonics}. In non-equilibrium, ${\rm Im}G_\bot$ is positive at small frequencies, and the gap does not appear. At large level spacing, the effects of plasmon gain and spectrum renormalization are relevant just in a narrow vicinity of the threshold of intraband tunneling. An increase in the temperature leads to the further narrowing of the 'gain window'. However, even at room temperature and relatively large bias ($V = 0.3$ V) there exists a range of frequencies corresponding to the net SP gain (solid part of orange line in Fig.~\ref{F4}B).

Finally, we briefly address the effects of rotational twist of graphene layers manifesting itself in the relative shift of the Dirac cones by the vectors $\Delta {\bf q}_{i}$ in the reciprocal space ($i=1...6$, the displacement vector for each of six pairs of Dirac cones is rotated by $\pi/3$ with respect to the next one). Neglecting the emerging small off-diagonal elements of the $\cal T$-matrix, one can prove that the tunnel  conductivity in the presence of twist $G^T_\bot({\bf q},\omega)$ is related to the tunnel conductivity of the aligned layers via
\begin{equation}
    G^T_\bot({\bf q},\omega) = \frac{1}{6}\sum\limits_{i=1}^{6}{G_\bot({\bf q}+\Delta {\bf q}_{i},\omega)}.
\end{equation}
In the presence of twist, the locus of collinear scattering singularities on the $\omega - q$ plane breaks down into six hyperbolas (or less, for some particular angles between ${\bf q}$ and $\Delta{\bf q}$). The acoustic plasmon dispersion develops an anticrossing with each of hyperbolas, demonstrating several frequency ranges with negative group velocity and gain. An example of the spectral function for SP propagating along $\Delta {\bf q}$ in one pair of valleys is shown in Fig.~\ref{F5} for $\hbar|\Delta {\bf q}|v_0 = 18$ meV (twist angle $\theta_T = 0.57^\circ$). In this example, there exist four curves corresponding to the singular plasmon gain and four for the singular absorption. Remarkably, the plasmon gain in twisted layers for certain directions of propagation can be greater than that in aligned layers, because the tunnel resonances can come closer to the unperturbed SP dispersion. Generally, the spectrum and gain of plasmons in twisted layers becomes anisotropic with six-fold rotational symmetry. 

The experimental observation of coherent plasmon amplification in coupled graphene layers poses strong constraints on the tunnel transparency and quality of the barrier layers. At the same time, the spontaneous emission of SPs upon tunneling~\cite{Lambe_PRL,Novotny_hBN} is readily observable for a wide class of dielectrics. The tunneling SP emission with their subsequent conversion into the free-space electromagnetic modes upon scattering might explain the observed terahertz electroluminescence from graphene-hBN-graphene diodes~\cite{THz_emission_DGL_experim}. The presence of luminescence in Ref.~\cite{THz_emission_DGL_experim} correlates with the presence of NDR in the static $I(V)$-curve, which supports the tunneling origin of the emission. The photon-assisted tunneling~\cite{Ryzhii_DGL_laser} may also contribute to the observed emission, however, the emission of photons carrying zero momentum is suppressed in samples with even a slight interlayer twist.

In conclusion, we have theoretically demonstrated a number of unique properties of surface plasmons in tunnel-coupled voltage-biased graphene layers, including the amplified propagation due to the resonant tunneling under interlayer population inversion, and a strong renormalization of dispersion law. The pronounced effect of tunneling on both spectrum and damping of plasmons results from singularities in the tunnel conductivity which are, in turn, inherited from the linear bands of graphene. Our findings can set the basis for novel active plasmonic devices based on van der Waals heterostructures, including compact plasmon sources and spasers.

The work of DS was supported by the grant \# 14-07-31315 of the Russian Foundation of Basic Research. The work at RIEC was supported by the Japan Society for Promotion of Science (Grant-in-Aid for Specially Promoted Research No. 23000008). The authors are grateful to V. Vyurkov, S. Fillipov, A. Dubinov, A. Arsenin and D. Fedyanin for helpful discussions.

\appendix
\setcounter{equation}{0}
\renewcommand{\theequation} {A\arabic{equation}}

\setcounter{figure}{0}
\renewcommand{\thefigure} {A\arabic{figure}}
\section{Supporting information}
\subsection{I. Plasmon modes supported by the double layer}
The plasmon spectra are obtained by a self-consistent solution of the Poisson's equation 
\begin{multline}
\label{Poisson}
-q^2 \delta \varphi(z) + \frac{\partial^2 \delta \varphi (z)}{\partial z^2} = \\
-\frac{4\pi}{\kappa}\left[ \delta Q_{t}\delta(z-d/2) + \delta Q_{b}\delta(z+d/2) \right],
\end{multline}
the continuity equations
\begin{equation}
-i\omega\delta Q_{t,b}+ i {\bf q}{\delta {\bf j}_{t,b}}= \mp \delta J_{\rm tun},
\end{equation}
and the linear-response relation between current density and electric field, $\delta {\bf j}_{t,b}= \sigma_\parallel({\bf q},\omega) \delta{\bf E}_{t,b}$, $\delta J_{\rm tun} = G_\bot({\bf q},\omega) (\delta\varphi_{t} - \delta\varphi_{b} )$. Here ${\bf q}$ is the two-dimensional plasmon wave vector, $d$ is the distance between layers, $\kappa$ is the background dielectric permittivity, $\delta Q_{t}$ and $\delta Q_{b}$ are the small-signal variations of charge density in the top and bottom layers, respectively, $\sigma_\parallel$ and $G_\bot$ are the in-plane and tunnel conductivities (note that the dimensionalities of these quantities are different), the indices $t$ and $b$ distinguish between the quantities corresponding to the top and bottom layers. In the absence of built-in voltage, due to the electron-hole symmetry, the charge densities in the layers are equal in modulus an opposite in sign, moreover, the layer conductivities are equal. This allows us to seek for the solutions of Eq.~(\ref{Poisson}) being symmetric and anti-symmetric with respect to $z$. A straightforward calculation brings us to the following dispersions~\cite{Voltage_controlled,Hwang_PRB_2GL}
\begin{equation}
\label{AntiSymmetric-disp}
1+\frac{2\pi i q}{\omega \kappa}\left[\sigma_\parallel ({\bf q},\omega) + \frac{2G_\bot ({\bf q},\omega )}{q^2}\right]\left( 1-e^{-qd} \right)=0
\end{equation}
for the antisymmetric (acoustic) mode, and
\begin{equation}
\label{Symmetric-disp}
1+\frac{2\pi i q}{\omega \kappa}\sigma_\parallel ({\bf q},\omega) \left( 1+e^{-qd} \right)
\end{equation}
for the symmetric (optical mode).

\begin{figure}[ht]
\includegraphics[width=0.9\linewidth]{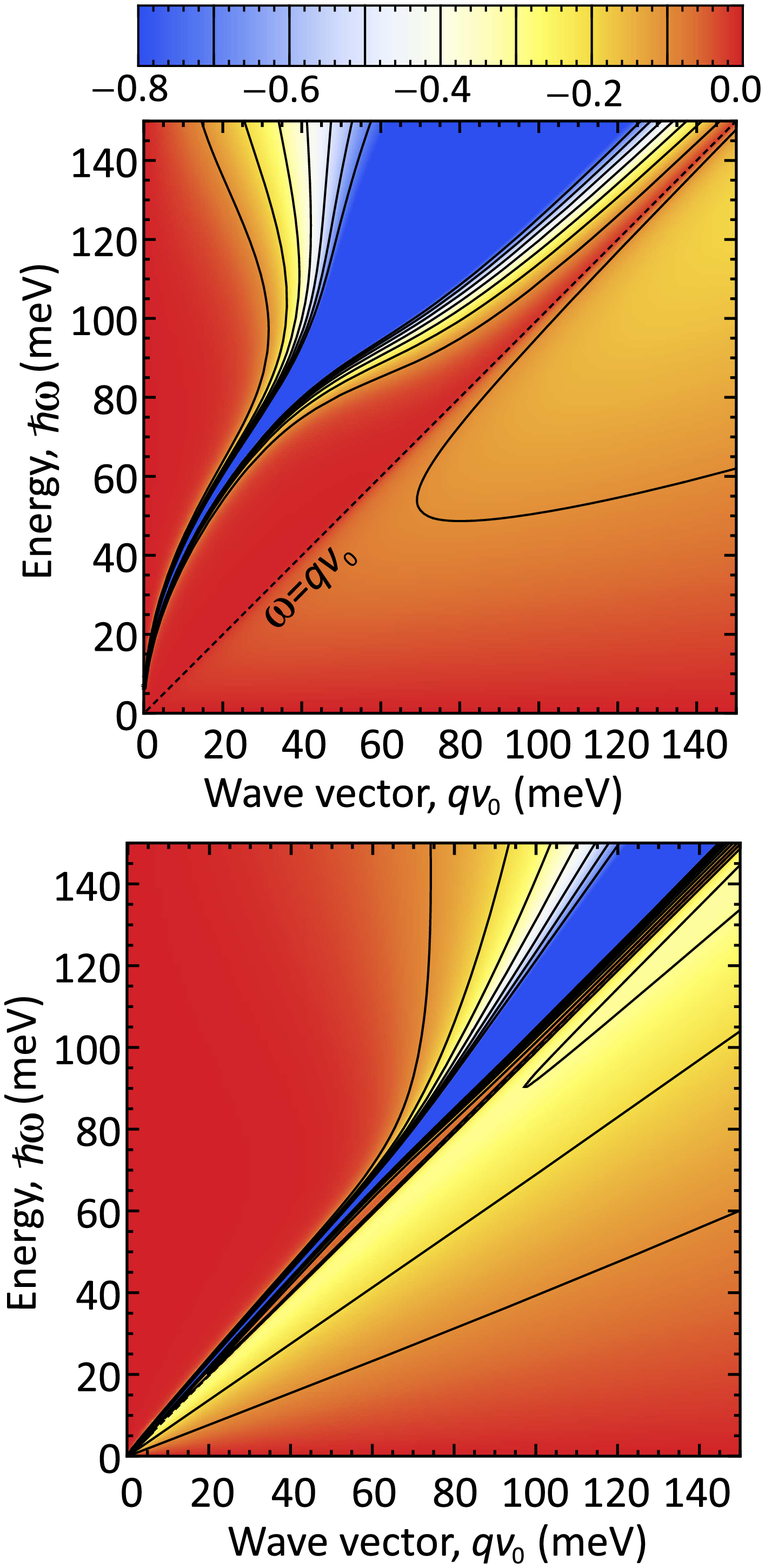}
\caption{
\label{PS} 
Spectra of acoustic and optical plasmons supported by the double graphene layer calculated for the following parameters: Fermi energy $\varepsilon_F = 100$ meV, temperature $T=300$ K, insulator thickness $d=3$ nm, dielectric constant $\kappa = 5$.
}
\end{figure}

Both equations (\ref{AntiSymmetric-disp}) and (\ref{Symmetric-disp}) can be considered as the zeros of the generalized polarizability of the double layer structure:
\begin{multline}
    \tilde\epsilon({\bf q},\omega) = \left\{1+\frac{2\pi i q}{\omega \kappa}\left[\sigma_\parallel ({\bf q},\omega) + \frac{2G_\bot ({\bf q},\omega )}{q^2}\right]\left( 1-e^{-qd} \right)\right\}\\
    \left\{1+\frac{2\pi i q}{\omega \kappa}\sigma_\parallel ({\bf q},\omega) \left( 1+e^{-qd} \right)\right\}.
\end{multline}

The imaginary part of the generalized polarizability inverted is the spectral function of the surface plasmons, 
\begin{equation}
    {\cal S}({\bf q},\omega) = {\rm Im}\tilde\epsilon^{-1}({\bf q},\omega),
\end{equation}
the positions of its peaks determine the SP spectra, its sign determines whether the excitations are amplified or damped, and the width of the peaks determines the magnitude of plasmon damping or gain. As the generalized polarizability decouples into the two terms with zeros yielding the dispersions of acoustic and optical modes, the spectral function  ${\cal S}({\bf q},\omega)$ can be also presented as a product of acoustic and optical plasmons' spectral functions:
\begin{equation}
    {\cal S}({\bf q},\omega) = {\cal S}_{\rm ac}({\bf q},\omega){\cal S}_{\rm opt}({\bf q},\omega).
\end{equation}

The spectral functions of acoustic and optical SPs are depicted in Fig.~\ref{PS} for highly doped ($\varepsilon_F=100$ meV) closely located graphene layers ($d=3$ nm).

It is possible to write down the analytical approximations to the plasmon spectra in the absence of tunneling. Being interested in the long-wavelength limit, $q d \ll 1$, we perform the expansions $1-e^{-qd}\approx qd$, $1+e^{-qd} \approx 2$. In the long-wavelength limit, the conductivity is essentially classical, moreover, the interband transitions do not affect the low-energy part of the spectra. With these assumptions, we use the following (collisionless) approximation for the conductivity which follows from the solution of the kinetic equation:
\begin{equation}
{{\sigma }_{\bf{q}\omega }}=i g \frac{{e^2}}{\hbar }\frac{{\tilde{\varepsilon}}_F}{2\pi \hbar}\frac{\omega }{q^2v_0^2}\left[ \frac{\omega }{\sqrt{{\omega^2}- q^2 v_0^2}}-1 \right],
\end{equation}
where $\tilde\varepsilon_F = T \ln(1+e^{\varepsilon_F/T})$. Equation~(\ref{AntiSymmetric-disp}) admits an analytical solution $\omega(q)$ with a sound-like dispersion
\begin{equation}
\omega_- = v_0 \frac{1 + 4 \alpha_c q_F d}{\sqrt{1 + 8 \alpha_c q_F d}} q.
\end{equation}
Here, we have introduced the Fermi wave vector $q_F = \tilde\varepsilon_F/\hbar v_0$, and the coupling constant $\alpha_C = e^2/\hbar\kappa v_0$. The velocity of the acoustic mode always exceeds the Fermi velocity, thus the Landau damping is avoided. The dispersion equation for the optical mode $\omega_+(q)$ is cubic, however, in the long-wavelength limit the spatial dispersion of conductivity can be neglected as the phase velocity of this mode significantly exceeds the Fermi velocity. The approximate relation for $\omega_+(q)$ has the following form
\begin{equation}
\omega_+ \approx v_0 \sqrt{4 \alpha_c q q_F}.
\end{equation}

\begin{figure}[ht]
\includegraphics[width=0.9\linewidth]{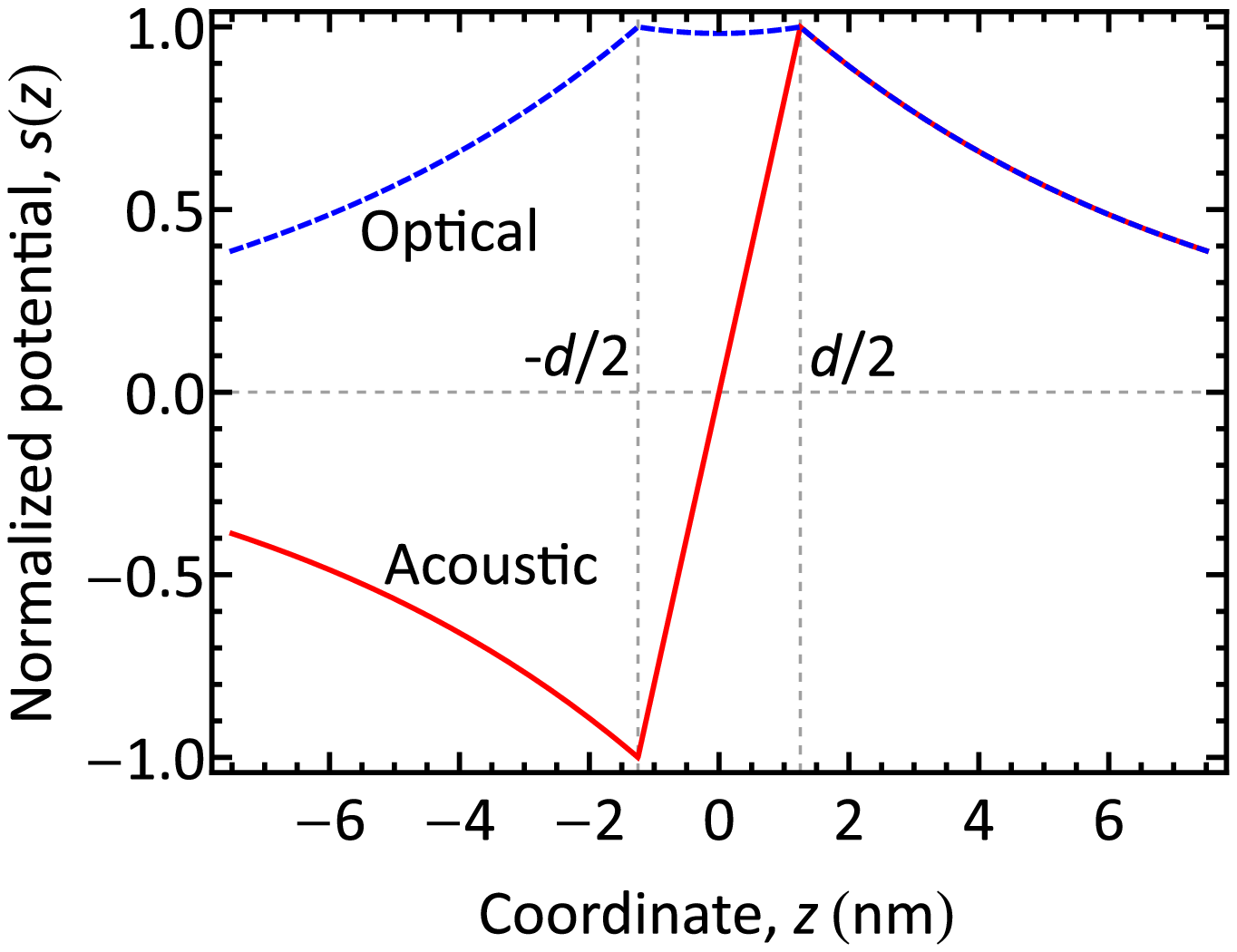}
\caption{\label{Shape} Dimensionless electric potential (normalized by its on-plane value) in the acoustic and optical modes calculated for the double layer structure with $d=2.5$ nm and wave vector $qv_0 = 100$ meV.
}
\end{figure}

In the subsequent calculations we shall also require the spatial dependence of the plasmon potential in the acoustic mode, which can be obtained from (\ref{Poisson}). It is convenient to present it as 
\begin{equation}
\label{Plasm-potential}
\delta\varphi(z) = \delta\varphi_0 s(z),
\end{equation}
where $\varphi_0$ is the electric potential on the top layer, and $s(z)$ is the dimensionless 'shape function' having the following form
\begin{equation}
\label{Shape-function}
s\left( z \right)=\left\{ \begin{aligned}
  & {{e}^{-q\left( z+d/2 \right)}},\,\,z<-d/2, \\ 
 & -\frac{\sinh \left( qz \right)}{\sinh \left( qd/2 \right)},\,\,\left| z \right|<d/2, \\ 
 & -{{e}^{-q\left( z-d/2 \right)}},\,\,z>d/2. \\ 
\end{aligned} \right.
\end{equation}
The spatial dependence of the shape functions for acoustic and optical modes is shown in Fig.~\ref{Shape}.

\subsection{II. Electron states in tunnel-coupled layers}
The tight-binding Hamiltonian of the tunnel-coupled graphene layers in the absence of the propagating plasmon [${\hat H}_0$ in Eq.~(\ref{Hamiltonian})] constitutes the blocks describing isolated graphene layers ${\hat H}_{G\pm}$ and the block describing tunnel hopping $\hat{\mathcal T}$. Such description of electron states is common for graphene bilayer with possible interlayer twist~\cite{Twisted_GBL}. In more comprehensive theories, the $\hat{\mathcal T}$-matrix is affected by the band structure of dielectric layer~\cite{Brey_PRA}. Here, for the sake of analytical traceability, we choose the tunneling matrix in its simplest form which is applicable to the AA-stacked perfectly aligned graphene bilayer, $\hat{\mathcal T} = \Omega \hat I$, where $\Omega$ can be interpreted as the tunnel hopping frequency. 

\begin{figure}[ht]
\includegraphics[width=0.9\linewidth]{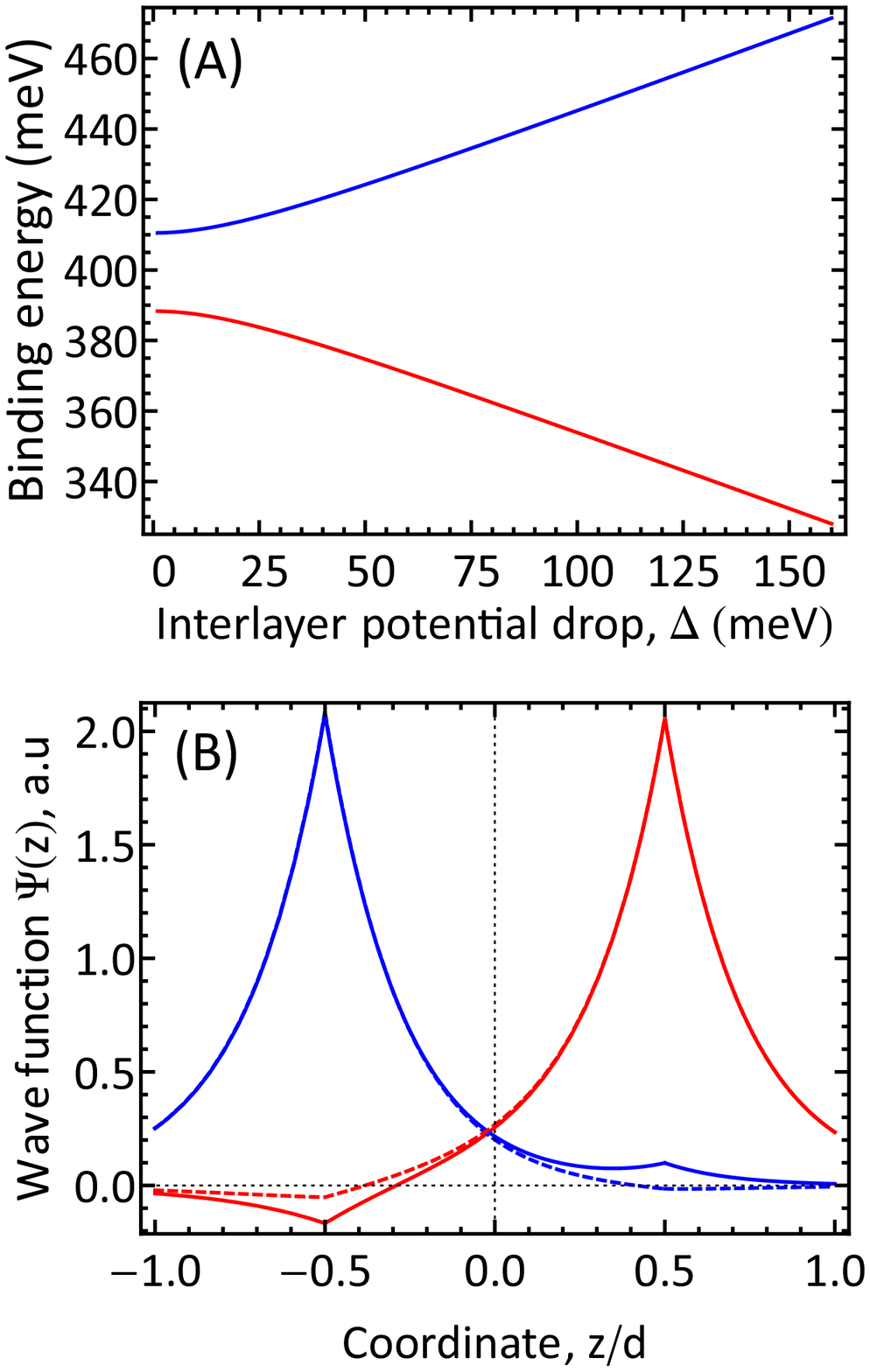}
\caption{\label{WF} Energy levels (A) and wave functions (B) of the tunnel-coupled graphene layers calculated for $\Delta=200$ meV and $2.5$ nm WS$_2$ as a tunnel barrier. Solid lines in (B) show the wave functions corresponding to $l=+1$ (red) and $l=-1$ (blue), while the dashed lines show the wave functions of the top and bottom layers obtained as a linear combination (\ref{LK}) of the eigen functions.
}
\end{figure}

To estimate its value, we switch for a while from the tight binding to the continuum description of electron states in the $z$-direction. We model each graphene layer with a delta-well~\cite{SP-lasing} 
\begin{equation}
    U_{t,b}(z) = 2\sqrt{\frac{\hbar^2 U_b}{2m^*}} \delta (z - z_{t,b}),
\end{equation}
 where the potential strength chosen to provide a correct value of electron work function $U_b$ from graphene to the surrounding dielectric, and $m^*$ is the effective electron mass in the dielectric. The effective Schrodinger equation in the presence of voltage bias $\Delta/e$ between graphene layers takes on the following form
\begin{multline}
-\frac{\hbar^2}{2m^*}\frac{\partial^2\Psi(z)}{\partial z^2} + \left[ U_t(z) + U_b(z) + U_F(z)\right]\Psi(z) = E \Psi(z),
\end{multline}
where $U_F$ is the potential energy created by the applied field
\begin{equation}
U_F\left( z \right)=\frac{\Delta}{2} \left\{ \begin{aligned}
  & 1,\,\,z<-d/2, \\ 
 & 2 z/d,\,\,\left| z \right|<d/2, \\ 
 & -1,\,\,z>d/2. \\ 
\end{aligned} \right.
\end{equation}
The solutions of effective Schrodinger equation represent decaying exponents at $|z| > d/2$, and a linear combination of Airy functions in the middle region $|z| < d/2$
\begin{equation}
\Psi_M(z) = C {\rm Ai} \left(-z/a + \varepsilon \right) + D {\rm Bi} \left(-z/a + \varepsilon \right),
\end{equation}
where $\varepsilon= 2m^* |E| a^2/\hbar^2 $ is the dimensionless energy and $a = (\hbar^2 d/ 2 m^*\Delta)^{1/3}$ is the effecive length in the electric field. A straightforward matching of the wave functions at the graphene layers yields the dispersion equation
\begin{widetext}
\begin{equation}
\label{Energy_spectrum}
\det \left( \begin{matrix}
   {e^{-k_1 d/2}} & -\text{Ai}\left( d/2a+\varepsilon  \right) & -\text{Bi}\left( d/2a+\varepsilon  \right) & 0  \\
   \left(2 k_b - k_1 \right){{e}^{-{k_1}d/2}} & -\frac{1}{a}\text{Ai}'\left( d/2a+\varepsilon  \right) & -\frac{1}{a}\text{Bi}'\left( d/2a+\varepsilon  \right) & 0  \\
   0 & -\text{Ai}\left( -d/2a+\varepsilon  \right) & -\text{Bi}\left( -d/2a+\varepsilon  \right) & {{e}^{-{{k}_{2}}d/2}}  \\
   0 & -\frac{1}{a}\text{Ai}'\left( -d/2a+\varepsilon  \right) & -\frac{1}{a}\text{Bi}'\left( -d/2a+\varepsilon  \right) & \left( 2k_b - k_2 \right){e^{-k_2 d/2}}  \\
\end{matrix} \right)=0,
\end{equation}
\end{widetext}
where $k_b = \sqrt{2m^* U_b/\hbar^2}$ is the decay constant of the bound state wave function in a single delta-well, $k_1 =\sqrt{2m^* (E + \Delta/2)/\hbar^2}$, $k_2 =\sqrt{2m^* (E - \Delta/2)/\hbar^2}$. Equation (\ref{Energy_spectrum}) yields two energy levels $E_l$ ($l = \pm 1$) which can be found only numerically (see Fig.~\ref{WF}A). The respective wave functions are shown in Fig.~\ref{WF}B, at strong bias they are {\it almost} the wave functions localized on the different layers (see the discussion below). Despite the complexity of Eq.~(\ref{Energy_spectrum}), the dependence of $E_l$ on the energy separation between layers $\Delta$ can be accurately modelled by
\begin{equation}
\label{Energy_spectrum_typical}
E_l (\Delta) = -U_b + \frac{l}{2} \sqrt{\left(E_{+1,\Delta=0} - E_{-1,\Delta=0}\right)^2 + \Delta^2}.
\end{equation}
The energy spectrum (\ref{Energy_spectrum_typical}) is typical for the tunnel coupled quantum wells~\cite{vasko_book}; the same functional dependence of energy levels on $\Delta$ is naturally obtained by diagonalizing the block Hamiltonian (\ref{Hamiltonian}),
\begin{equation}
\label{Energy_spectrum_block}
E_l (\Delta) = -U_b + l \sqrt{\Omega^2+ \frac{\Delta^2}{4}}.
\end{equation}
This allows us to estimate the tunnel coupling $\Omega$ as half the energy splitting of states in double graphene layer well in the absence of applied bias
\begin{equation}
\Omega=\frac{1}{2}\left[E_{+1,\Delta=0} - E_{-1,\Delta=0} \right].
\end{equation}

The $l$-index governs the $z$-localization of electron in a biased double quantum well. At large bias $\Delta \gg \Omega$, the delta-wells interact weakly, thus $l = +1$ corresponds to the state localized almost completely in the top layer and $l = -1$ corresponds to the electron in the bottom layer. The wave functions corresponding to a relatively strong bias $\Delta = 200$ meV are shown in Fig.~\ref{WF}. It is simple to relate the true eigen functions $\Psi_+(z)$ and $\Psi_-(z)$ to the functions located on the top and bottom layers $\Psi_t(z)$ and $\Psi_b(z)$
\begin{gather}
\label{LK}
    \Psi_b = \cos\alpha \Psi_- + \sin\alpha \Psi_+,\\
    \Psi_t = -\sin\alpha \Psi_- + \cos\alpha \Psi_+,
\end{gather}
where
\begin{equation}
    \cos\alpha = \frac{2\Omega}{\sqrt{(2\Omega)^2 + (\Delta - \tilde\Delta)^2}}.
\end{equation}

At small bias $\Delta \ll \Omega$ the wave function of $l = +1$ is odd and that of $l = -1$ is even.

\subsection{III. Electron-plasmon interaction and solution of the quantum Liouville equation}

The presence of plasmon propagating along the double graphene layer results in an additional potential energy of electron
\begin{equation}
\label{Plasm_energy}
\delta \hat V ({\bf r},t) = e \delta\varphi(z) e^{i(q x-\omega t)},
\end{equation}
where we assume the direction of plasmon propagation to be along the $x$-axis, and the dependence of potential on the $z$-coordinate is given by Eqs.~(\ref{Plasm-potential}) and (\ref{Shape-function}). The additional terms in Hamiltonian due to the vector-potential are negligible as far as the speed of light substantially exceeds the plasmon velocity.

\begin{figure}[ht]
\includegraphics[width=0.9\linewidth]{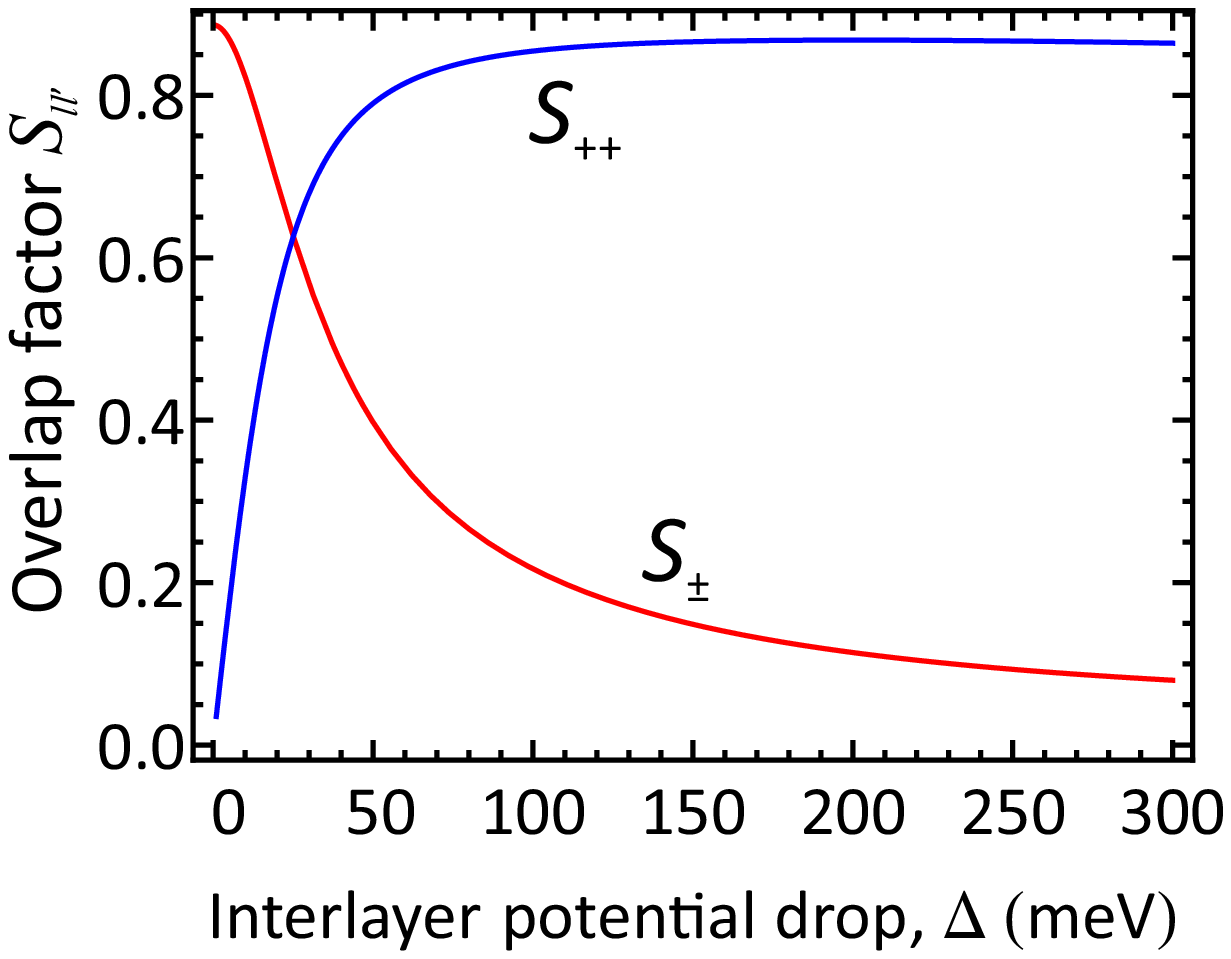}
\caption{\label{OF} Dependence of the overlap factors $S_{++}$ and $S_\pm$ of $\hat H_0$-eigenfunctions and dimensionless plasmon potential $s(z)$ calculated for the WS$_2$ (2.5 nm) dielectric layer.
}
\end{figure}

With our choice of the tight-binding basis functions as those localized on a definite layer and on a definite lattice cite, we shall require 16 matrix elements of the potential energy (\ref{Plasm_energy}) connecting those basis states. However, it is more convenient to work out the matrix elements of (\ref{Plasm_energy}) connecting the {\it eigen} states of Hamiltonian (\ref{Hamiltonian}). The good quantum numbers of these states are the in-plane momentum ${\bf p}$, the band index $s = \pm 1$  ($+1$ for the conduction band and $-1$ for the valence band) and the $l$ - index discussed above. The respective matrix elements are
\begin{multline}
\bra{{\bf p}sl}\delta \hat V \ket{{\bf p}'s'l'} = \\
\delta_{{\bf p},{\bf p}'-{\bf q}} u^{ss'}_{{\bf pp}'} e \delta\varphi_0 \int_{-\infty}^{\infty}{\Psi^*_l(z) s(z) \Psi_{l'}(z)}.
\end{multline}

We introduce the shorthand notations for the overlap factors of dimensionless plasmon potential and eigen functions of coupled layers
\begin{gather}
    S_{++} = \int_{-\infty}^{\infty}{\Psi^*_{+1}(z) s(z) \Psi_{+1}(z)},\\
    S_{\pm} = \int_{-\infty}^{\infty}{\Psi^*_{+1}(z) s(z) \Psi_{-1}(z)},
\end{gather}
and, obviously, $S_{--} = -S_{++}$, $S_{\pm} = S_{\mp}$.

The dependence of the overlap factors $S_{++}$ and $S_{\pm}$ on the interlayer potential drop $\Delta$ is shown in Fig.~\ref{OF}. We note that these overlap factors weakly depend on the plasmon wave vector $q$ as far as it is much smaller than electron wave function decay constant $k_b$. In this approximation, one can set $s(z)\approx 2z/d$ for $|z|<d/2$, $s(z)\approx 1$ at $z < - d/2$, and $s(z)\approx 1$ at $z > d/2$. 

Having obtained the matrix elements of electron-plasmon interaction, we pass to the solution of the quantum Liuoville equation for the electron density matrix $\hat \rho$. In the linear response, the latter is decomposed as $\hat \rho = \hat \rho^{(0)} + \delta \hat\rho$, where $\delta \hat\rho$ emerges due to the plasmon field. This component of the density matrix is found from
\begin{equation}
\label{Neumann}
i\hbar\frac{\partial \delta \hat\rho}{\partial t} = [{\hat H}_0, \delta \hat\rho] + [\delta \hat V,\hat\rho^{(0)}].
\end{equation}

Considering the harmonic time dependence, Eq.~(\ref{Neumann}) is exactly (non-perturbatively) solved in the diagonal basis of $\hat H_0$. In this basis, the commutator 
\begin{equation}
    [{\hat H}_0,\delta \hat\rho]_{\alpha \beta} = (\varepsilon_\alpha - \varepsilon_\beta)\delta \rho_{\alpha \beta},
\end{equation} 
where $\alpha$ and $\beta$ run over good quantum numbers ${\bf p}$, $s$ and $l$. Thus, one readily writes down the solution
\begin{equation}
\delta\rho_{\alpha\beta} = \frac{\left[\delta \hat V , \hat\rho^{(0)} \right]_{\alpha\beta}}{\hbar\omega + i\delta - (\varepsilon_\alpha - \varepsilon_\beta) }.
\end{equation}

The first-order correction $\delta \hat\rho$ is now expressed through the density matrix in the absence of plasmon field $\hat\rho^{(0)}$. A particular choice of $\hat\rho^{(0)}$ requires the solution of kinetic equation in the voltage-biased tunnel-coupled layers, however, in several limiting cases the situation is greatly simplified~\cite{Kazarinov_Suris}. If the tunneling rate $\Omega$ is slower than the electron energy relaxation rate $\nu_{\varepsilon}$ (e.g., due to phonons and carrier-carrier scattering), the quasi-equilibrium distribution function is established in each individual layer. In this situation, $\hat\rho^{(0)}$ is diagonal in the basis formed by the wave functions localized on top and bottom layers its elements being the respective Fermi distribution functions. In the other limiting case, when tunneling is stronger than scattering ($\Omega \gg \nu_\varepsilon$), the electron is 'collectivized' by the two layers, and the density matrix $\hat\rho^{(0)}$ is approximately diagonal in the basis of ${\hat H}_0$-eigenstates. For the parameters used in our calculations, $\hbar\Omega\approx 10$ meV exceeds the relaxation rate $\hbar \nu\approx 1$ meV, and the latter limiting case is justified. Setting $\rho^{(0)}_{\alpha\beta} = f_\alpha \delta_{\alpha\beta}$, where $f$ is the Fermi distribution function, we find
\begin{multline}
\bra{{\bf p},s,l}{\delta \hat \rho}\ket{{\bf p}'s'l'} = \\
= \delta\varphi_0 S_{ll'} u^{ss'}_{{\bf pp}'} \frac{f^{s'l'}_{{\bf p}'} - f^{sl}_{\bf p}}{\hbar \omega + i\delta - (\varepsilon^{sl}_{\bf p} - \varepsilon^{s'l'}_{{\bf p}'}) }.
\end{multline}
We note that a different choice of the zero-order density matrix also leads to the emergence of the negative tunnel conductivity, with a larger coefficient in front of $G_\bot$.

The subsequent calculation of the in-plane and tunnel conductivities is based on the following relations. From the charge conservation on the top layer one has
\begin{multline}
\label{Continuity-2}
    \frac{\partial \delta Q_t}{\partial t} = - {\bf \nabla} {\delta \bf j} - \delta J_{\rm tun} = \\
    q^2\left[\sigma_\parallel({\bf q},\omega) + 2\frac{G_\bot({\bf q},\omega)}{q^2}\right]\delta\varphi_0.
\end{multline}
On the other hand, the time derivative of the charge density can be obtained by statistical averaging of the operator 
\begin{equation}
    \frac{\partial Q_{\alpha\beta}}{\partial t} = \frac{i}{\hbar} Q_{\alpha \beta} (\varepsilon_\alpha - \varepsilon_\beta),
\end{equation}
where, as before, the indices $\alpha$ and $\beta$ run over in-plane momentum $\bf p$, band index $s$, and $z$-localization index $l$. The rule of statistical averaging of $\partial\hat Q/\partial t$ in extended form reads
\begin{multline}
\label{Averaging}
    \frac{\partial \delta Q_t}{\partial t} = {\rm Tr}\frac{\partial \hat Q_t}{\partial t}\delta \hat\rho=\\
    -\frac{i}{\hbar}\sum\limits_{{\bf p}ss'll'}{\bra{{\bf p}_+ s' l' }\hat Q\ket{{\bf p}_- s l}\bra{{\bf p}_- s l} \delta\hat\rho \ket{{\bf p}_+ s' l'} ( \varepsilon^{sl}_{\bf p_-} - \varepsilon^{s'l'}_{\bf p_+} ) }.
\end{multline}

Due to the linear dependence of $\delta\hat\rho$ on the plasmon potential amplitude $\delta\varphi_0$, the average time derivative of the charge density in Eq.~(\ref{Averaging}) is also a linear function of $\delta\varphi_0$. The proportionality coefficient, according to Eq.~(\ref{Continuity-2}), is the sought-for combination of conductivities $q^2 \sigma_\parallel + 2 G_\bot$. The terms in the sum, Eq.~(\ref{Averaging}), with non-equal $l$-indices are related to the tunnel conductivity, and those with equal $l$-indices -- to the in-plane conductivity. 

Actually, the distinction between in-plane and tunnel conductivity is meaningful only in the case of weak coupling (or strong bias $\Delta \gg \Omega$). In the this case $S_{++} \cos\theta_M \rightarrow 1 $, and Eq.~(\ref{In-plane}) yields the conductivity of a single graphene layer~\cite{Falkovsky-Varlamov}. In the same limit, the tunnel conductivity, Eq.~(\ref{Out-plane}) possesses a small prefactor $S_{\pm} \sin\theta_M \propto e^{-2 k_b d}$, where $k_b$ is the decay constant of the electron wave function. In the opposite case of weak bias $\Omega \gtrsim \Delta$, the notions of in-plane and tunnel conductivities lose their meaning as the states of individual layers are highly mixed~\cite{DasSarma-PRL-tunnel-plasmon}. Ultimately, at zero bias, $\sigma_\parallel$ vanishes, which reflects the impossibility of electron transitions between states with the same $z$-symmetry under the perturbation odd in $z$. 

However, even in the case of strong bias $\Delta \gg \Omega$, the in-plane conductivity of tunnel-coupled layers responding to the plasmon field is renormalized compared to its value for a single isolated layer in uniform field $\sigma_0$, namely
\begin{equation}
    \sigma_\parallel = S_{++} \cos\theta_M \sigma_0.
\end{equation}
The factor $S_{++}<1$ comes from the broadening of the electron cloud beyond a single layer and non-uniformity of the plasmon field. Loosely speaking, a part of the electron wave function feels the reduced magnitude of the plasmon field ${\cal E}_\parallel$ outside of graphene layer. The factor $\cos\theta_M < 1$ comes from the mixing of electron states in individual layers forming the state with definite value of $l$.

\subsection{IV. Analytical approximations to the in-plane and tunnel conductivity}

Despite a complex structure of Eqs.~(\ref{In-plane}) and (\ref{Out-plane}), several analytical approximations can be made in the frequency range of interest $\hbar \omega < 2\varepsilon_F$, where the plasmons are weakly damped -- at least, for the real part of conductivity that determines absorption or gain. For brevity, in this section we work with 'god-given units' $\hbar = v_0 \equiv 1 $ We start with the evaluation of in-plane interband conductivity associated with the electron transitions from the valence band to the conduction band
\begin{multline}
  {\rm Re} {\sigma_0}^{v\rightarrow c}( {\bf q},\omega )=-i g \frac{e^2}{\omega}\times\\
\sum\limits_{\bf{p}}{ {| {\bf v}_{{\bf p}{\bf p}'}^{vc} |^2} [f_{{\bf p}_-}^v-f_{{\bf p}_-}^c] \delta(\omega - \varepsilon_{\bf p_+} - \varepsilon_{\bf p_-})}.    
\end{multline}
Here ${\bf v}^{\rm{vc}}_{ {\bf p}+,{\bf p}-}$ is the interband matrix element of velocity operator in graphene $\hat{\bf v} =  {\boldsymbol \sigma}$, and $\epsilon_{\bf p} = p$ is the dispersion law. Known the eigen functions of graphene Hamiltonian $\hat{H}_G = {\boldsymbol \sigma } {\bf p}$,
\begin{equation}
\ket{ s {\bf p}} =\frac{1}{\sqrt{2}}\left( 
\begin{aligned}
  & {{e}^{-i{{\theta }_{\bf{p}}}/2}} \\ 
 & {s{e}^{i{{\theta }_{\bf{p}}}/2}} \\ 
\end{aligned} \right) e^{i {\bf p r}},
\end{equation}
one readily finds $\bra{c {\bf p}_-} \hat{v}_x \ket{v {\bf p}_+} = i \sin\left[(\theta_{{\bf p}+} + \theta_{{\bf p}-})/2\right]$. The subsequent calculations are conveniently performed in the elliptic coordinates
\begin{equation}
{\bf p} =\frac{q}{2}\left\{ \cosh u\cos v, \sinh u \sin v \right\}.
\end{equation}
In these coordinates $|{\bf p}_\pm| = (q/2)[\cosh u \pm \cos v]$, $|\bra{c {\bf p}_-} \hat{v}_x \ket{v {\bf p}_+}|^2 dp_x dp_y = q^2/4 \cosh^2u\sin^2v du dv$. This leads us to
\begin{multline}
\label{Inter-exact}
{\rm{Re}}\sigma^{v\rightarrow c}_0({\bf q},\omega) = \frac{e^2}{2\pi}\frac{\omega}{\sqrt{\omega^2 - q^2}}\int\limits_0^\pi dv \sin^2v \times \\
\left\{f_0\left[-\frac{\omega}{2} + \frac{q}{2} \cos v \right] - f_0\left[\frac{\omega}{2} + \frac{q}{2} \cos v \right] \right\}.
\end{multline}

To proceed further, we note that in the domain of interest $\omega > q$ one always has $ q \cos v < \omega$. Due to this fact, the difference of distribution functions is a smooth function of $v$, while the prefactor $\sin^2v$ varies strongly. This allows us to integrate $\sin^2v$ exactly, and replace the difference of distribution functions with its angular average. This leads us to
\begin{multline}
\label{Inter-approx2}
{\rm{Re}}\sigma^{v \rightarrow c}_{0}({\bf q},\omega) \approx \frac{e^2}{4} \frac{T \omega}{q} \chi(q,\omega) \times\\
\ln\frac{\cosh \frac{\varepsilon_F}{T} + \cosh \frac{\omega + q}{2T}}{\cosh \frac{\varepsilon_F}{T} + \cosh \frac{\omega - q}{2T}},
\end{multline}
where we have introduced a resonant factor 
\begin{equation}
    \chi(q,\omega) = \frac{\theta(\omega)}{{\sqrt{\omega^2 - q^2}}}.
\end{equation}

Clearly, the neglect of spatial dispersion in the case of acoustic SPs with velocity slightly exceeding the Fermi velocity results in an underestimation of the real part of the interband conductivity and, hence, of the damping.

Similar approximations can be made to evaluate the interlayer interband conductivity, the only difference is that electrons in different layers have different chemical potentials. We present these results without derivation
\begin{widetext}
\begin{gather}
    \frac{2G_\bot^{v \rightarrow c}}{q^2} = -e^2 \frac{T\omega}{2 q}
    \left\{\chi(q,\tilde\Delta - \omega) \ln\displaystyle{\frac{\cosh\frac{q + eV-\omega}{4T}}{\cosh\frac{q - eV + \omega}{4T}}} - \chi(q,\tilde\Delta + \omega) \ln\displaystyle{\frac{\cosh\frac{q + eV + \omega}{4T}}{\cosh\frac{q - eV - \omega}{4T}}} \right\},\\
    \frac{2G_\bot^{c \rightarrow v}}{q^2} = -e^2\frac{T\omega}{2 q} \
    \left\{\chi(q, \omega - \tilde\Delta) 
    \ln\displaystyle{\frac{\cosh\frac{q + eV - \omega}{4T}}{\cosh\frac{q - eV +\omega}{4T}}} - \chi(q,- \tilde\Delta - \omega) \ln\displaystyle{\frac{\cosh\frac{q + eV + \omega}{4T}}{\cosh\frac{q - eV -\omega}{4T}}} \right\}.
\end{gather}
\end{widetext}

\begin{figure}[ht]
\includegraphics[width=0.9\linewidth]{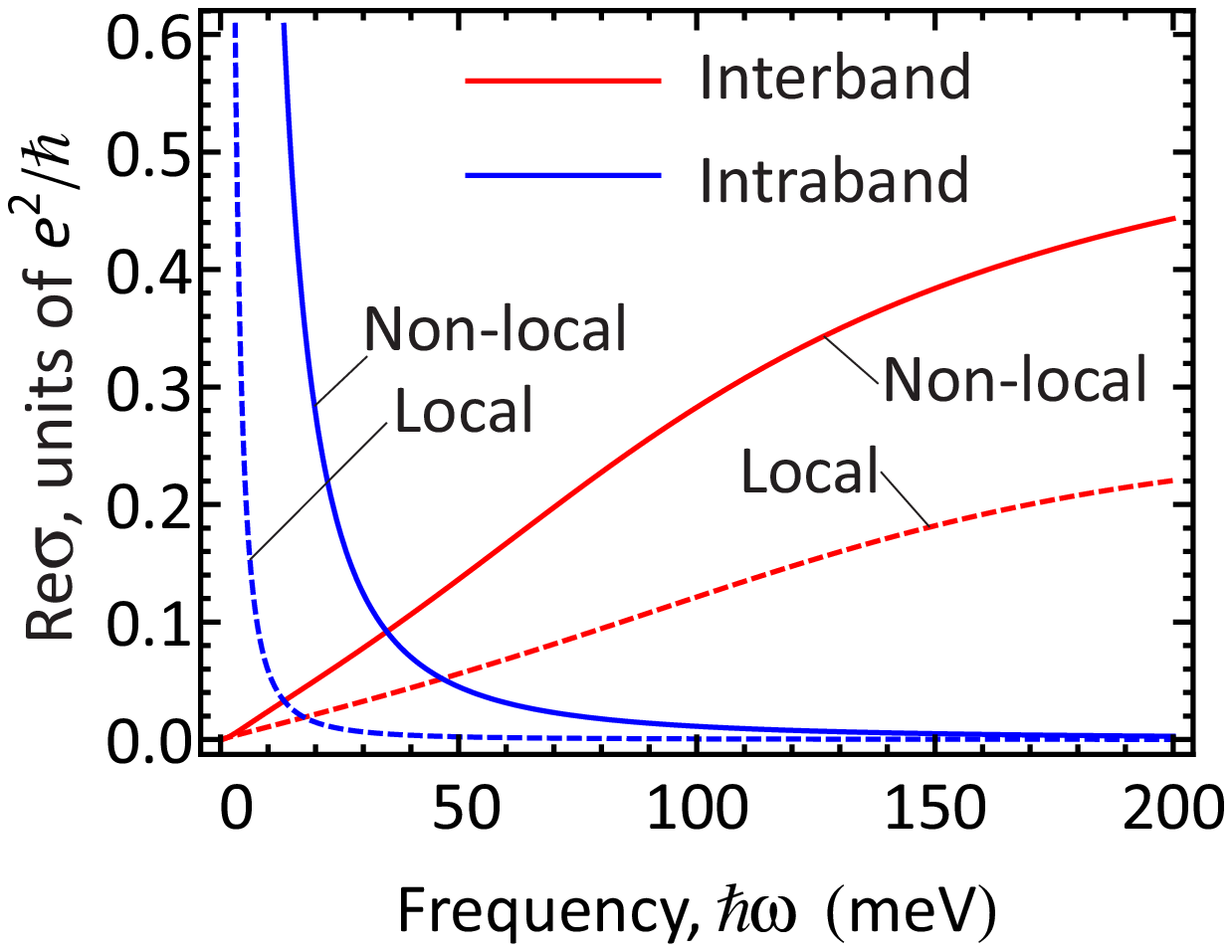}
\caption{\label{Net_Sigma} Comparison of the real parts of the interband (red) and intraband (blue) conductivities of a single graphene layer evaluated in the local limit (dashed) and at finite wave vector corresponding to the acoustic SP dispersion $q = \omega /s$ (solid). The parameters used in the calculation are $\varepsilon_F = 100$ meV, $T=300$ K, $s=1.2v_0$. Acoustic phonons are considered as the main carrier relaxation mechanism.   
}
\end{figure}

We now pass to the in-plane conductivity associated with the intraband transitions. Here, we can restrict ourselves to the classical description of the electron motion justified at frequencies $\omega \ll \varepsilon_F$, $q \ll q_F$ -- otherwise, strong interband SP damping takes place. Clearly, one could work out the terms with $s=s'$ and $l=l'$ in Eq.~(\ref{In-plane}), however, the accurate inclusion of carrier scattering in such equations is challenging. Instead, we use the kinetic equation to evaluate $\sigma_\parallel^{c\rightarrow c}$; this formalism allows an inclusion of carrier scattering in a consistent manner. One should, however, keep in mind that in non-local case $q\neq 0$ a simple $\tau_p$-approximation is not particle-conserving. A particle-conserving account of collisions is achieved with the Bhatnagar-Gross-Krook collision integral~\cite{BGK-collisions} in the right-hand side of the kinetic equation,
\begin{multline}
\label{Kinetic-BGK}
-i\omega \delta f ({\bf p}) +i{\bf q v} \delta f ( {\bf p} ) + i e {\bf q v}{\delta\varphi}\frac{\partial {f_0}}{\partial \varepsilon }=\\
-\nu \left[ \delta f ( {\bf p} ) + \frac{d{{\varepsilon }_{F}}}{dn}\frac{\partial {f_0}}{\partial \varepsilon }\delta {n} \right].
\end{multline}
Here $\delta f (\bf{p})$ is the sought-for field-dependent correction to the equilibrium electron distribution function $f_0$, $\delta n_{\bf q}$ is the respective correction to the electron density, ${\bf v} = {\bf p}/p$ is the quasi-particle velocity, and $\nu$ is the electron collision frequency which is assumed to be energy-independent. The current density, associated with the distribution function $\delta f (\bf{p})$ reads:
\begin{equation}
\label{Current_BGK}
\delta {\bf j} = - e g \sum_{\bf p}{
{\bf v} \frac{df_0}{d\varepsilon} 
\frac{i e {\bf qv} \delta\varphi - i \nu (d\varepsilon_F/dn)\delta n }{\omega + i \nu - {\bf q v}} 
}.
\end{equation}
Recalling the relation between small-signal variations of density and current, $\omega \delta n= q \delta {\bf j}$, and evaluating the integrals in Eq.~(\ref{Current_BGK}), we find the in-plane intraband conductivity:
\begin{equation}
\label{Sigma_intra}
\sigma_{\rm{intra}}({\bf q},\omega) = \frac{i g e^2 \tilde\varepsilon_F}{(2\pi)^2q}\displaystyle{\frac{J_2(\frac{\omega + i \nu}{q})}{1 - \frac{i\nu}{2\pi\omega}J_1(\frac{\omega + i \nu}{q})}},
\end{equation}
where 
\begin{equation}
J_n(x) = \int_0^{2\pi}{\frac{\cos^n\theta d\theta}{x - \cos\theta}}.
\end{equation}
Similar to the real part of the interband absorption, the intraband absorption is generally larger in the non-local case $q\neq 0$ compared to the local case. This difference is illustrated in Fig.~(\ref{Net_Sigma}), where the local ($q=0$) and non-local expressions at the acoustic plasmon dispersion ($q = \omega /s$) are compared. This result is in agreement with the recent measurements of plasmon propagation length in graphene on hBN: the local Drude formula underestimated the plasmon damping, and the account of non-locality was crucial to explain the experimental data~\cite{Principi_plasmon_loss_hBN}.

Finally, we provide the analytical estimates for the intraband tunnel conductivity, which mainly governs the tunneling effects on the plasmon dispersion. After passing to the elliptic coordinates in terms with $l\neq l'$ and $s=s'$ one readily finds
\begin{multline}
    {\rm Re}\frac{2G^{c\rightarrow c}_\bot}{q^2} = - e^2 S_{\pm}\sin\theta_M\frac{\omega}{2\pi}\\
    \left\{\psi(q,\omega - \tilde\Delta) I \left(\frac{q}{2T},\frac{eV-\omega}{2T}\right) - \psi(q,\omega + \tilde\Delta) I \left(\frac{q}{2T},\frac{eV + \omega}{2T}\right) \right\},
\end{multline}
where we have introduced the resonant factor associated with the intraband interlayer transitions
\begin{equation}
    \psi(q,\omega) = \frac{1}{\sqrt{q^2 - \omega^2}},
\end{equation}
and an auxiliary dimensionless integral
\begin{equation}
    I(\alpha,\beta) = \int\limits_1^\infty{dt\sqrt{t^2 - 1}\left[F(\alpha t - \beta) - F(\alpha t+\beta)\right]},
\end{equation}
here $F(\zeta) =(1+e^\zeta)^{-1}$ is the dimensionless Fermi function.

The collinear tunneling singularities are smeared in the presence of carrier scattering. To account for the latter, we replace the delta-peaked spectral functions of individual particles in the expressions for conductivity (\ref{In-plane}) and (\ref{Out-plane}) with Lorentz functions using the following rule
\begin{multline}
\label{Broadening}
    \sum_{\bf p}{\frac{1}{\omega+i\delta - (\varepsilon^{sl}_{\bf p} - \varepsilon^{s'l'}_{\bf p'})}} = \\
    \int{d\varepsilon d\varepsilon' \sum_{\bf p}{\frac{ \delta(\varepsilon - \varepsilon^{sl}_{\bf p}) \delta(\varepsilon' - \varepsilon^{s'l'}_{\bf p'}) }{\omega + i \delta - (\varepsilon - \varepsilon')}}} \Rightarrow  \\
    \frac{1}{(2\pi)^2}
    \int{
        d\varepsilon d\varepsilon' 
                \sum_{\bf p}{
                    \frac
                        {{\cal A}_{sl}({\bf p},\varepsilon) {\cal A}_{s'l'}({\bf p}',\varepsilon) }
                        {\omega + i \delta - (\varepsilon - \varepsilon')}
                        }
        }.
\end{multline}

The spectral function is given by
\begin{equation}
   {\cal A}_{sl}({\bf p},\varepsilon) = \frac{2 \Sigma''_{sl}({\bf p}, \varepsilon)}{[\varepsilon - \varepsilon^{sl}_{\bf p}]^2 + [\Sigma''_{sl}({\bf p}, \varepsilon)]^2}, 
\end{equation}
where we have taken the imaginary part of the spectral function as half the electron-phonon collision frequency evaluated at the Fermi surface~\cite{Vasko-Ryzhii}
\begin{equation}
    2\Sigma''_{sl}({\bf p}, \varepsilon) = \left.\frac{\varepsilon}{T}\frac{D^2 T^2}{2\rho s^2 v_0^2}\right|_{\varepsilon = \varepsilon_F}.
\end{equation}
The approximation (\ref{Broadening}) corresponds to the neglect of vertex corrections in the current-current correlator represented by the bubble diagram. Moreover, the interlayer electron-phonon interactions are neglected. While these assumptions can be hardly justified, they do not affect much the calculated plasmon spectral functions and dispersions because the plasmon spectra do not enter the domain of singular conductivity. Nevertheless, the account of scattering in a more consistent manner may lead to the new results. As example, Kazarinov and Suris have shown that the interference of scattering events in different layers can lead to a sufficient decrease in the effective collision frequency governing the width of resonances in dynamic tunnel conductivity~\cite{Kazarinov_Suris}. This effective collision frequency should be much less than transport collision frequency and, a fortiori, the relaxation frequency. The extension of their results to the case of plasmon-assisted tunneling will be the subject of the future work.

\bibliography{Bibliography}

\begin{thebibliography}{41}%
\makeatletter
\providecommand \@ifxundefined [1]{%
 \@ifx{#1\undefined}
}%
\providecommand \@ifnum [1]{%
 \ifnum #1\expandafter \@firstoftwo
 \else \expandafter \@secondoftwo
 \fi
}%
\providecommand \@ifx [1]{%
 \ifx #1\expandafter \@firstoftwo
 \else \expandafter \@secondoftwo
 \fi
}%
\providecommand \natexlab [1]{#1}%
\providecommand \enquote  [1]{``#1''}%
\providecommand \bibnamefont  [1]{#1}%
\providecommand \bibfnamefont [1]{#1}%
\providecommand \citenamefont [1]{#1}%
\providecommand \href@noop [0]{\@secondoftwo}%
\providecommand \href [0]{\begingroup \@sanitize@url \@href}%
\providecommand \@href[1]{\@@startlink{#1}\@@href}%
\providecommand \@@href[1]{\endgroup#1\@@endlink}%
\providecommand \@sanitize@url [0]{\catcode `\\12\catcode `\$12\catcode
  `\&12\catcode `\#12\catcode `\^12\catcode `\_12\catcode `\%12\relax}%
\providecommand \@@startlink[1]{}%
\providecommand \@@endlink[0]{}%
\providecommand \url  [0]{\begingroup\@sanitize@url \@url }%
\providecommand \@url [1]{\endgroup\@href {#1}{\urlprefix }}%
\providecommand \urlprefix  [0]{URL }%
\providecommand \Eprint [0]{\href }%
\providecommand \doibase [0]{http://dx.doi.org/}%
\providecommand \selectlanguage [0]{\@gobble}%
\providecommand \bibinfo  [0]{\@secondoftwo}%
\providecommand \bibfield  [0]{\@secondoftwo}%
\providecommand \translation [1]{[#1]}%
\providecommand \BibitemOpen [0]{}%
\providecommand \bibitemStop [0]{}%
\providecommand \bibitemNoStop [0]{.\EOS\space}%
\providecommand \EOS [0]{\spacefactor3000\relax}%
\providecommand \BibitemShut  [1]{\csname bibitem#1\endcsname}%
\let\auto@bib@innerbib\@empty
%</preamble>
\bibitem [{\citenamefont {Grigorenko}\ \emph {et~al.}(2012)\citenamefont
  {Grigorenko}, \citenamefont {Polini},\ and\ \citenamefont
  {Novoselov}}]{Graphene_plasmonics-1}%
  \BibitemOpen
  \bibfield  {author} {\bibinfo {author} {\bibfnamefont {A.}~\bibnamefont
  {Grigorenko}}, \bibinfo {author} {\bibfnamefont {M.}~\bibnamefont {Polini}},
  \ and\ \bibinfo {author} {\bibfnamefont {K.}~\bibnamefont {Novoselov}},\
  }\href@noop {} {\bibfield  {journal} {\bibinfo  {journal} {Nat. Photonics}\
  }\textbf {\bibinfo {volume} {6}},\ \bibinfo {pages} {749} (\bibinfo {year}
  {2012})}\BibitemShut {NoStop}%
\bibitem [{\citenamefont {Koppens}\ \emph {et~al.}(2011)\citenamefont
  {Koppens}, \citenamefont {Chang},\ and\ \citenamefont {Garcia~de
  Abajo}}]{Graphene_plasmonics-2}%
  \BibitemOpen
  \bibfield  {author} {\bibinfo {author} {\bibfnamefont {F.~H.~L.}\
  \bibnamefont {Koppens}}, \bibinfo {author} {\bibfnamefont {D.~E.}\
  \bibnamefont {Chang}}, \ and\ \bibinfo {author} {\bibfnamefont {F.~J.}\
  \bibnamefont {Garcia~de Abajo}},\ }\href {\doibase 10.1021/nl201771h}
  {\bibfield  {journal} {\bibinfo  {journal} {Nano Lett.}\ }\textbf {\bibinfo
  {volume} {11}},\ \bibinfo {pages} {3370} (\bibinfo {year}
  {2011})}\BibitemShut {NoStop}%
\bibitem [{\citenamefont {Hwang}\ and\ \citenamefont
  {Das~Sarma}(2007)}]{Das_Sarma_Plasmons}%
  \BibitemOpen
  \bibfield  {author} {\bibinfo {author} {\bibfnamefont {E.~H.}\ \bibnamefont
  {Hwang}}\ and\ \bibinfo {author} {\bibfnamefont {S.}~\bibnamefont
  {Das~Sarma}},\ }\href {\doibase 10.1103/PhysRevB.75.205418} {\bibfield
  {journal} {\bibinfo  {journal} {Phys. Rev. B}\ }\textbf {\bibinfo {volume}
  {75}},\ \bibinfo {pages} {205418} (\bibinfo {year} {2007})}\BibitemShut
  {NoStop}%
\bibitem [{\citenamefont {Ryzhii}\ \emph {et~al.}(2007)\citenamefont {Ryzhii},
  \citenamefont {Satou},\ and\ \citenamefont {Otsuji}}]{Ryzhii-plasmons}%
  \BibitemOpen
  \bibfield  {author} {\bibinfo {author} {\bibfnamefont {V.}~\bibnamefont
  {Ryzhii}}, \bibinfo {author} {\bibfnamefont {A.}~\bibnamefont {Satou}}, \
  and\ \bibinfo {author} {\bibfnamefont {T.}~\bibnamefont {Otsuji}},\ }\href
  {\doibase 10.1063/1.2426904} {\bibfield  {journal} {\bibinfo  {journal} {J.
  Appl. Phys.}\ }\textbf {\bibinfo {volume} {101}},\ \bibinfo {eid} {024509}
  (\bibinfo {year} {2007})}\BibitemShut {NoStop}%
\bibitem [{\citenamefont {Mikhailov}\ and\ \citenamefont
  {Ziegler}(2007)}]{Mikhailov_new_mode}%
  \BibitemOpen
  \bibfield  {author} {\bibinfo {author} {\bibfnamefont {S.~A.}\ \bibnamefont
  {Mikhailov}}\ and\ \bibinfo {author} {\bibfnamefont {K.}~\bibnamefont
  {Ziegler}},\ }\href {\doibase 10.1103/PhysRevLett.99.016803} {\bibfield
  {journal} {\bibinfo  {journal} {Phys. Rev. Lett.}\ }\textbf {\bibinfo
  {volume} {99}},\ \bibinfo {pages} {016803} (\bibinfo {year}
  {2007})}\BibitemShut {NoStop}%
\bibitem [{\citenamefont {Svintsov}\ \emph {et~al.}(2012)\citenamefont
  {Svintsov}, \citenamefont {Vyurkov}, \citenamefont {Yurchenko}, \citenamefont
  {Otsuji},\ and\ \citenamefont {Ryzhii}}]{Our-hydrodynamic}%
  \BibitemOpen
  \bibfield  {author} {\bibinfo {author} {\bibfnamefont {D.}~\bibnamefont
  {Svintsov}}, \bibinfo {author} {\bibfnamefont {V.}~\bibnamefont {Vyurkov}},
  \bibinfo {author} {\bibfnamefont {S.}~\bibnamefont {Yurchenko}}, \bibinfo
  {author} {\bibfnamefont {T.}~\bibnamefont {Otsuji}}, \ and\ \bibinfo {author}
  {\bibfnamefont {V.}~\bibnamefont {Ryzhii}},\ }\href {\doibase
  10.1063/1.4705382} {\bibfield  {journal} {\bibinfo  {journal} {J. Appl.
  Phys.}\ }\textbf {\bibinfo {volume} {111}},\ \bibinfo {eid} {083715}
  (\bibinfo {year} {2012})}\BibitemShut {NoStop}%
\bibitem [{\citenamefont {Gangadharaiah}\ \emph {et~al.}(2008)\citenamefont
  {Gangadharaiah}, \citenamefont {Farid},\ and\ \citenamefont
  {Mishchenko}}]{New_plasmon_mode}%
  \BibitemOpen
  \bibfield  {author} {\bibinfo {author} {\bibfnamefont {S.}~\bibnamefont
  {Gangadharaiah}}, \bibinfo {author} {\bibfnamefont {A.~M.}\ \bibnamefont
  {Farid}}, \ and\ \bibinfo {author} {\bibfnamefont {E.~G.}\ \bibnamefont
  {Mishchenko}},\ }\href {\doibase 10.1103/PhysRevLett.100.166802} {\bibfield
  {journal} {\bibinfo  {journal} {Phys. Rev. Lett.}\ }\textbf {\bibinfo
  {volume} {100}},\ \bibinfo {pages} {166802} (\bibinfo {year}
  {2008})}\BibitemShut {NoStop}%
\bibitem [{\citenamefont {Woessner}\ \emph {et~al.}(2014)\citenamefont
  {Woessner}, \citenamefont {Lundeberg}, \citenamefont {Gao}, \citenamefont
  {Principi}, \citenamefont {Alonso-Gonz{\'a}lez}, \citenamefont {Carrega},
  \citenamefont {Watanabe}, \citenamefont {Taniguchi}, \citenamefont {Vignale},
  \citenamefont {Polini}, \citenamefont {Hone}, \citenamefont {Hillenbrand},\
  and\ \citenamefont {Koppens}}]{Koppens_nano_imaging_hBN}%
  \BibitemOpen
  \bibfield  {author} {\bibinfo {author} {\bibfnamefont {A.}~\bibnamefont
  {Woessner}}, \bibinfo {author} {\bibfnamefont {M.~B.}\ \bibnamefont
  {Lundeberg}}, \bibinfo {author} {\bibfnamefont {Y.}~\bibnamefont {Gao}},
  \bibinfo {author} {\bibfnamefont {A.}~\bibnamefont {Principi}}, \bibinfo
  {author} {\bibfnamefont {P.}~\bibnamefont {Alonso-Gonz{\'a}lez}}, \bibinfo
  {author} {\bibfnamefont {M.}~\bibnamefont {Carrega}}, \bibinfo {author}
  {\bibfnamefont {K.}~\bibnamefont {Watanabe}}, \bibinfo {author}
  {\bibfnamefont {T.}~\bibnamefont {Taniguchi}}, \bibinfo {author}
  {\bibfnamefont {G.}~\bibnamefont {Vignale}}, \bibinfo {author} {\bibfnamefont
  {M.}~\bibnamefont {Polini}}, \bibinfo {author} {\bibfnamefont
  {J.}~\bibnamefont {Hone}}, \bibinfo {author} {\bibfnamefont {R.}~\bibnamefont
  {Hillenbrand}}, \ and\ \bibinfo {author} {\bibfnamefont {F.~H.~L.}\
  \bibnamefont {Koppens}},\ }\href@noop {} {\bibfield  {journal} {\bibinfo
  {journal} {Nat. Mater.}\ } (\bibinfo {year} {2014})}\BibitemShut {NoStop}%
\bibitem [{\citenamefont {Tomadin}\ and\ \citenamefont
  {Polini}(2013)}]{Polini_FET}%
  \BibitemOpen
  \bibfield  {author} {\bibinfo {author} {\bibfnamefont {A.}~\bibnamefont
  {Tomadin}}\ and\ \bibinfo {author} {\bibfnamefont {M.}~\bibnamefont
  {Polini}},\ }\href {\doibase 10.1103/PhysRevB.88.205426} {\bibfield
  {journal} {\bibinfo  {journal} {Phys. Rev. B}\ }\textbf {\bibinfo {volume}
  {88}},\ \bibinfo {pages} {205426} (\bibinfo {year} {2013})}\BibitemShut
  {NoStop}%
\bibitem [{\citenamefont {Svintsov}\ \emph
  {et~al.}(2013{\natexlab{a}})\citenamefont {Svintsov}, \citenamefont
  {Vyurkov}, \citenamefont {Ryzhii},\ and\ \citenamefont {Otsuji}}]{Our_NLHD}%
  \BibitemOpen
  \bibfield  {author} {\bibinfo {author} {\bibfnamefont {D.}~\bibnamefont
  {Svintsov}}, \bibinfo {author} {\bibfnamefont {V.}~\bibnamefont {Vyurkov}},
  \bibinfo {author} {\bibfnamefont {V.}~\bibnamefont {Ryzhii}}, \ and\ \bibinfo
  {author} {\bibfnamefont {T.}~\bibnamefont {Otsuji}},\ }\href {\doibase
  10.1103/PhysRevB.88.245444} {\bibfield  {journal} {\bibinfo  {journal} {Phys.
  Rev. B}\ }\textbf {\bibinfo {volume} {88}},\ \bibinfo {pages} {245444}
  (\bibinfo {year} {2013}{\natexlab{a}})}\BibitemShut {NoStop}%
\bibitem [{\citenamefont {Dubinov}\ \emph {et~al.}(2011)\citenamefont
  {Dubinov}, \citenamefont {Aleshkin}, \citenamefont {Mitin}, \citenamefont
  {Otsuji},\ and\ \citenamefont {Ryzhii}}]{Dubinov_JPCM}%
  \BibitemOpen
  \bibfield  {author} {\bibinfo {author} {\bibfnamefont {A.~A.}\ \bibnamefont
  {Dubinov}}, \bibinfo {author} {\bibfnamefont {V.~Y.}\ \bibnamefont
  {Aleshkin}}, \bibinfo {author} {\bibfnamefont {V.}~\bibnamefont {Mitin}},
  \bibinfo {author} {\bibfnamefont {T.}~\bibnamefont {Otsuji}}, \ and\ \bibinfo
  {author} {\bibfnamefont {V.}~\bibnamefont {Ryzhii}},\ }\href
  {http://stacks.iop.org/0953-8984/23/i=14/a=145302} {\bibfield  {journal}
  {\bibinfo  {journal} {J. Phys.: Cond. Mat.}\ }\textbf {\bibinfo {volume}
  {23}},\ \bibinfo {pages} {145302} (\bibinfo {year} {2011})}\BibitemShut
  {NoStop}%
\bibitem [{\citenamefont {Rana}(2008)}]{Rana_IEEE}%
  \BibitemOpen
  \bibfield  {author} {\bibinfo {author} {\bibfnamefont {F.}~\bibnamefont
  {Rana}},\ }\href {\doibase 10.1109/TNANO.2007.910334} {\bibfield  {journal}
  {\bibinfo  {journal} {IEEE T. Nanotechnol.}\ }\textbf {\bibinfo {volume}
  {7}},\ \bibinfo {pages} {91} (\bibinfo {year} {2008})}\BibitemShut {NoStop}%
\bibitem [{\citenamefont {Feiginov}\ and\ \citenamefont
  {Volkov}(1998)}]{Feiginov_Volkov}%
  \BibitemOpen
  \bibfield  {author} {\bibinfo {author} {\bibfnamefont {M.}~\bibnamefont
  {Feiginov}}\ and\ \bibinfo {author} {\bibfnamefont {V.}~\bibnamefont
  {Volkov}},\ }\href@noop {} {\bibfield  {journal} {\bibinfo  {journal} {JETP
  Letters}\ }\textbf {\bibinfo {volume} {68}},\ \bibinfo {pages} {662}
  (\bibinfo {year} {1998})}\BibitemShut {NoStop}%
\bibitem [{\citenamefont {Ryzhii}\ and\ \citenamefont
  {Shur}(2001)}]{Ryzhii_Shur_JJAP}%
  \BibitemOpen
  \bibfield  {author} {\bibinfo {author} {\bibfnamefont {V.}~\bibnamefont
  {Ryzhii}}\ and\ \bibinfo {author} {\bibfnamefont {M.}~\bibnamefont {Shur}},\
  }\href {http://stacks.iop.org/1347-4065/40/i=2R/a=546} {\bibfield  {journal}
  {\bibinfo  {journal} {Jpn. J. Appl. Phys.}\ }\textbf {\bibinfo {volume}
  {40}},\ \bibinfo {pages} {546} (\bibinfo {year} {2001})}\BibitemShut
  {NoStop}%
\bibitem [{\citenamefont {Sensale-Rodriguez}(2013)}]{Berardi_APL}%
  \BibitemOpen
  \bibfield  {author} {\bibinfo {author} {\bibfnamefont {B.}~\bibnamefont
  {Sensale-Rodriguez}},\ }\href {\doibase http://dx.doi.org/10.1063/1.4821221}
  {\bibfield  {journal} {\bibinfo  {journal} {Appl. Phys. Lett.}\ }\textbf
  {\bibinfo {volume} {103}},\ \bibinfo {eid} {123109} (\bibinfo {year}
  {2013})}\BibitemShut {NoStop}%
\bibitem [{\citenamefont {Ryzhii}\ \emph
  {et~al.}(2013{\natexlab{a}})\citenamefont {Ryzhii}, \citenamefont {Satou},
  \citenamefont {Otsuji}, \citenamefont {Ryzhii}, \citenamefont {Mitin},\ and\
  \citenamefont {Shur}}]{Dynamic_effects}%
  \BibitemOpen
  \bibfield  {author} {\bibinfo {author} {\bibfnamefont {V.}~\bibnamefont
  {Ryzhii}}, \bibinfo {author} {\bibfnamefont {A.}~\bibnamefont {Satou}},
  \bibinfo {author} {\bibfnamefont {T.}~\bibnamefont {Otsuji}}, \bibinfo
  {author} {\bibfnamefont {M.}~\bibnamefont {Ryzhii}}, \bibinfo {author}
  {\bibfnamefont {V.}~\bibnamefont {Mitin}}, \ and\ \bibinfo {author}
  {\bibfnamefont {M.~S.}\ \bibnamefont {Shur}},\ }\href
  {http://stacks.iop.org/0022-3727/46/i=31/a=315107} {\bibfield  {journal}
  {\bibinfo  {journal} {J. Phys. D: Appl. Phys.}\ }\textbf {\bibinfo {volume}
  {46}},\ \bibinfo {pages} {315107} (\bibinfo {year}
  {2013}{\natexlab{a}})}\BibitemShut {NoStop}%
\bibitem [{\citenamefont {Das~Sarma}\ and\ \citenamefont
  {Hwang}(1998)}]{DasSarma-PRL-tunnel-plasmon}%
  \BibitemOpen
  \bibfield  {author} {\bibinfo {author} {\bibfnamefont {S.}~\bibnamefont
  {Das~Sarma}}\ and\ \bibinfo {author} {\bibfnamefont {E.~H.}\ \bibnamefont
  {Hwang}},\ }\href {\doibase 10.1103/PhysRevLett.81.4216} {\bibfield
  {journal} {\bibinfo  {journal} {Phys. Rev. Lett.}\ }\textbf {\bibinfo
  {volume} {81}},\ \bibinfo {pages} {4216} (\bibinfo {year}
  {1998})}\BibitemShut {NoStop}%
\bibitem [{\citenamefont {Hwang}\ and\ \citenamefont
  {Das~Sarma}(2009)}]{Hwang_PRB_2GL}%
  \BibitemOpen
  \bibfield  {author} {\bibinfo {author} {\bibfnamefont {E.~H.}\ \bibnamefont
  {Hwang}}\ and\ \bibinfo {author} {\bibfnamefont {S.}~\bibnamefont
  {Das~Sarma}},\ }\href {\doibase 10.1103/PhysRevB.80.205405} {\bibfield
  {journal} {\bibinfo  {journal} {Phys. Rev. B}\ }\textbf {\bibinfo {volume}
  {80}},\ \bibinfo {pages} {205405} (\bibinfo {year} {2009})}\BibitemShut
  {NoStop}%
\bibitem [{\citenamefont {Svintsov}\ \emph
  {et~al.}(2013{\natexlab{b}})\citenamefont {Svintsov}, \citenamefont
  {Vyurkov}, \citenamefont {Ryzhii},\ and\ \citenamefont
  {Otsuji}}]{Voltage_controlled}%
  \BibitemOpen
  \bibfield  {author} {\bibinfo {author} {\bibfnamefont {D.}~\bibnamefont
  {Svintsov}}, \bibinfo {author} {\bibfnamefont {V.}~\bibnamefont {Vyurkov}},
  \bibinfo {author} {\bibfnamefont {V.}~\bibnamefont {Ryzhii}}, \ and\ \bibinfo
  {author} {\bibfnamefont {T.}~\bibnamefont {Otsuji}},\ }\href {\doibase
  http://dx.doi.org/10.1063/1.4789818} {\bibfield  {journal} {\bibinfo
  {journal} {J. Appl. Phys.}\ }\textbf {\bibinfo {volume} {113}},\ \bibinfo
  {eid} {053701} (\bibinfo {year} {2013}{\natexlab{b}})}\BibitemShut {NoStop}%
\bibitem [{\citenamefont {Brey}(2014)}]{Brey_PRA}%
  \BibitemOpen
  \bibfield  {author} {\bibinfo {author} {\bibfnamefont {L.}~\bibnamefont
  {Brey}},\ }\href {\doibase 10.1103/PhysRevApplied.2.014003} {\bibfield
  {journal} {\bibinfo  {journal} {Phys. Rev. Applied}\ }\textbf {\bibinfo
  {volume} {2}},\ \bibinfo {pages} {014003} (\bibinfo {year}
  {2014})}\BibitemShut {NoStop}%
\bibitem [{\citenamefont {Vasko}(2013)}]{Vasko_PRB}%
  \BibitemOpen
  \bibfield  {author} {\bibinfo {author} {\bibfnamefont {F.~T.}\ \bibnamefont
  {Vasko}},\ }\href {\doibase 10.1103/PhysRevB.87.075424} {\bibfield  {journal}
  {\bibinfo  {journal} {Phys. Rev. B}\ }\textbf {\bibinfo {volume} {87}},\
  \bibinfo {pages} {075424} (\bibinfo {year} {2013})}\BibitemShut {NoStop}%
\bibitem [{\citenamefont {Ryzhii}\ \emph
  {et~al.}(2013{\natexlab{b}})\citenamefont {Ryzhii}, \citenamefont {Dubinov},
  \citenamefont {Aleshkin}, \citenamefont {Ryzhii},\ and\ \citenamefont
  {Otsuji}}]{Ryzhii_DGL_laser}%
  \BibitemOpen
  \bibfield  {author} {\bibinfo {author} {\bibfnamefont {V.}~\bibnamefont
  {Ryzhii}}, \bibinfo {author} {\bibfnamefont {A.~A.}\ \bibnamefont {Dubinov}},
  \bibinfo {author} {\bibfnamefont {V.~Y.}\ \bibnamefont {Aleshkin}}, \bibinfo
  {author} {\bibfnamefont {M.}~\bibnamefont {Ryzhii}}, \ and\ \bibinfo {author}
  {\bibfnamefont {T.}~\bibnamefont {Otsuji}},\ }\href {\doibase
  http://dx.doi.org/10.1063/1.4826113} {\bibfield  {journal} {\bibinfo
  {journal} {Appl. Phys. Lett.}\ }\textbf {\bibinfo {volume} {103}},\ \bibinfo
  {eid} {163507} (\bibinfo {year} {2013}{\natexlab{b}})}\BibitemShut {NoStop}%
\bibitem [{\citenamefont {Bistritzer}\ and\ \citenamefont
  {MacDonald}(2011)}]{Twisted_GBL}%
  \BibitemOpen
  \bibfield  {author} {\bibinfo {author} {\bibfnamefont {R.}~\bibnamefont
  {Bistritzer}}\ and\ \bibinfo {author} {\bibfnamefont {A.~H.}\ \bibnamefont
  {MacDonald}},\ }\href@noop {} {\bibfield  {journal} {\bibinfo  {journal}
  {Proc. Nat. Acad. Sci.}\ }\textbf {\bibinfo {volume} {108}},\ \bibinfo
  {pages} {12233} (\bibinfo {year} {2011})}\BibitemShut {NoStop}%
\bibitem [{\citenamefont {Shi}\ \emph {et~al.}(2013)\citenamefont {Shi},
  \citenamefont {Pan}, \citenamefont {Zhang},\ and\ \citenamefont
  {Yakobson}}]{Band_parameters_MOS2}%
  \BibitemOpen
  \bibfield  {author} {\bibinfo {author} {\bibfnamefont {H.}~\bibnamefont
  {Shi}}, \bibinfo {author} {\bibfnamefont {H.}~\bibnamefont {Pan}}, \bibinfo
  {author} {\bibfnamefont {Y.-W.}\ \bibnamefont {Zhang}}, \ and\ \bibinfo
  {author} {\bibfnamefont {B.~I.}\ \bibnamefont {Yakobson}},\ }\href {\doibase
  10.1103/PhysRevB.87.155304} {\bibfield  {journal} {\bibinfo  {journal} {Phys.
  Rev. B}\ }\textbf {\bibinfo {volume} {87}},\ \bibinfo {pages} {155304}
  (\bibinfo {year} {2013})}\BibitemShut {NoStop}%
\bibitem [{\citenamefont {Vasko}\ and\ \citenamefont
  {Kuznetsov}(2012)}]{vasko_book}%
  \BibitemOpen
  \bibfield  {author} {\bibinfo {author} {\bibfnamefont {F.~T.}\ \bibnamefont
  {Vasko}}\ and\ \bibinfo {author} {\bibfnamefont {A.~V.}\ \bibnamefont
  {Kuznetsov}},\ }\href@noop {} {\emph {\bibinfo {title} {Electronic states and
  optical transitions in semiconductor heterostructures}}}\ (\bibinfo
  {publisher} {Springer Science \& Business Media},\ \bibinfo {year}
  {2012})\BibitemShut {NoStop}%
\bibitem [{Note1()}]{Note1}%
  \BibitemOpen
  \bibinfo {note} {Note that the in-plane conductivity is also renormalized due
  to the delocalization of electron wave function outside of graphene and
  nonuniformity of plasmon field, $S_{++}<1$}\BibitemShut {NoStop}%
\bibitem [{\citenamefont {Kazarinov}\ and\ \citenamefont
  {Suris}(1972)}]{Kazarinov_Suris}%
  \BibitemOpen
  \bibfield  {author} {\bibinfo {author} {\bibfnamefont {R.}~\bibnamefont
  {Kazarinov}}\ and\ \bibinfo {author} {\bibfnamefont {R.}~\bibnamefont
  {Suris}},\ }\href@noop {} {\bibfield  {journal} {\bibinfo  {journal} {Sov.
  Phys. Semicond.}\ }\textbf {\bibinfo {volume} {6}},\ \bibinfo {pages} {148}
  (\bibinfo {year} {1972})}\BibitemShut {NoStop}%
\bibitem [{\citenamefont {Faist}\ \emph {et~al.}(1994)\citenamefont {Faist},
  \citenamefont {Capasso}, \citenamefont {Sivco}, \citenamefont {Sirtori},
  \citenamefont {Hutchinson},\ and\ \citenamefont {Cho}}]{Capasso_Science}%
  \BibitemOpen
  \bibfield  {author} {\bibinfo {author} {\bibfnamefont {J.}~\bibnamefont
  {Faist}}, \bibinfo {author} {\bibfnamefont {F.}~\bibnamefont {Capasso}},
  \bibinfo {author} {\bibfnamefont {D.~L.}\ \bibnamefont {Sivco}}, \bibinfo
  {author} {\bibfnamefont {C.}~\bibnamefont {Sirtori}}, \bibinfo {author}
  {\bibfnamefont {A.~L.}\ \bibnamefont {Hutchinson}}, \ and\ \bibinfo {author}
  {\bibfnamefont {A.~Y.}\ \bibnamefont {Cho}},\ }\href@noop {} {\bibfield
  {journal} {\bibinfo  {journal} {Science}\ }\textbf {\bibinfo {volume}
  {264}},\ \bibinfo {pages} {553} (\bibinfo {year} {1994})}\BibitemShut
  {NoStop}%
\bibitem [{\citenamefont {Greenaway}\ \emph {et~al.}(2015)\citenamefont
  {Greenaway}, \citenamefont {Vdovin}, \citenamefont {Mishchenko},
  \citenamefont {Makarovsky}, \citenamefont {Patan{\`e}}, \citenamefont
  {Wallbank}, \citenamefont {Cao}, \citenamefont {Kretinin}, \citenamefont
  {Zhu}, \citenamefont {Morozov} \emph {et~al.}}]{Chiral_tunneling}%
  \BibitemOpen
  \bibfield  {author} {\bibinfo {author} {\bibfnamefont {M.}~\bibnamefont
  {Greenaway}}, \bibinfo {author} {\bibfnamefont {E.}~\bibnamefont {Vdovin}},
  \bibinfo {author} {\bibfnamefont {A.}~\bibnamefont {Mishchenko}}, \bibinfo
  {author} {\bibfnamefont {O.}~\bibnamefont {Makarovsky}}, \bibinfo {author}
  {\bibfnamefont {A.}~\bibnamefont {Patan{\`e}}}, \bibinfo {author}
  {\bibfnamefont {J.}~\bibnamefont {Wallbank}}, \bibinfo {author}
  {\bibfnamefont {Y.}~\bibnamefont {Cao}}, \bibinfo {author} {\bibfnamefont
  {A.}~\bibnamefont {Kretinin}}, \bibinfo {author} {\bibfnamefont
  {M.}~\bibnamefont {Zhu}}, \bibinfo {author} {\bibfnamefont {S.}~\bibnamefont
  {Morozov}},  \emph {et~al.},\ }\href@noop {} {\bibfield  {journal} {\bibinfo
  {journal} {Nature Physics}\ }\textbf {\bibinfo {volume} {11}},\ \bibinfo
  {pages} {1057} (\bibinfo {year} {2015})}\BibitemShut {NoStop}%
\bibitem [{Note2()}]{Note2}%
  \BibitemOpen
  \bibinfo {note} {Alternatively, the singularities in the tunnel conductivity
  can be traced back to the van Hove singularities in the joint density of
  states~\cite {Brey_PRA}}\BibitemShut {NoStop}%
\bibitem [{\citenamefont {Principi}\ \emph {et~al.}(2014)\citenamefont
  {Principi}, \citenamefont {Carrega}, \citenamefont {Lundeberg}, \citenamefont
  {Woessner}, \citenamefont {Koppens}, \citenamefont {Vignale},\ and\
  \citenamefont {Polini}}]{Principi_plasmon_loss_hBN}%
  \BibitemOpen
  \bibfield  {author} {\bibinfo {author} {\bibfnamefont {A.}~\bibnamefont
  {Principi}}, \bibinfo {author} {\bibfnamefont {M.}~\bibnamefont {Carrega}},
  \bibinfo {author} {\bibfnamefont {M.~B.}\ \bibnamefont {Lundeberg}}, \bibinfo
  {author} {\bibfnamefont {A.}~\bibnamefont {Woessner}}, \bibinfo {author}
  {\bibfnamefont {F.~H.~L.}\ \bibnamefont {Koppens}}, \bibinfo {author}
  {\bibfnamefont {G.}~\bibnamefont {Vignale}}, \ and\ \bibinfo {author}
  {\bibfnamefont {M.}~\bibnamefont {Polini}},\ }\href {\doibase
  10.1103/PhysRevB.90.165408} {\bibfield  {journal} {\bibinfo  {journal} {Phys.
  Rev. B}\ }\textbf {\bibinfo {volume} {90}},\ \bibinfo {pages} {165408}
  (\bibinfo {year} {2014})}\BibitemShut {NoStop}%
\bibitem [{\citenamefont {Vasko}\ and\ \citenamefont
  {Ryzhii}(2007)}]{Vasko-Ryzhii}%
  \BibitemOpen
  \bibfield  {author} {\bibinfo {author} {\bibfnamefont {F.~T.}\ \bibnamefont
  {Vasko}}\ and\ \bibinfo {author} {\bibfnamefont {V.}~\bibnamefont {Ryzhii}},\
  }\href {\doibase 10.1103/PhysRevB.76.233404} {\bibfield  {journal} {\bibinfo
  {journal} {Phys. Rev. B}\ }\textbf {\bibinfo {volume} {76}},\ \bibinfo
  {pages} {233404} (\bibinfo {year} {2007})}\BibitemShut {NoStop}%
\bibitem [{\citenamefont {Bolotin}\ \emph {et~al.}(2008)\citenamefont
  {Bolotin}, \citenamefont {Sikes}, \citenamefont {Hone}, \citenamefont
  {Stormer},\ and\ \citenamefont {Kim}}]{PRL_Bolotin}%
  \BibitemOpen
  \bibfield  {author} {\bibinfo {author} {\bibfnamefont {K.~I.}\ \bibnamefont
  {Bolotin}}, \bibinfo {author} {\bibfnamefont {K.~J.}\ \bibnamefont {Sikes}},
  \bibinfo {author} {\bibfnamefont {J.}~\bibnamefont {Hone}}, \bibinfo {author}
  {\bibfnamefont {H.~L.}\ \bibnamefont {Stormer}}, \ and\ \bibinfo {author}
  {\bibfnamefont {P.}~\bibnamefont {Kim}},\ }\href {\doibase
  10.1103/PhysRevLett.101.096802} {\bibfield  {journal} {\bibinfo  {journal}
  {Phys. Rev. Lett.}\ }\textbf {\bibinfo {volume} {101}},\ \bibinfo {pages}
  {096802} (\bibinfo {year} {2008})}\BibitemShut {NoStop}%
\bibitem [{\citenamefont {Principi}\ \emph {et~al.}(2011)\citenamefont
  {Principi}, \citenamefont {Asgari},\ and\ \citenamefont
  {Polini}}]{Polini-soundarons}%
  \BibitemOpen
  \bibfield  {author} {\bibinfo {author} {\bibfnamefont {A.}~\bibnamefont
  {Principi}}, \bibinfo {author} {\bibfnamefont {R.}~\bibnamefont {Asgari}}, \
  and\ \bibinfo {author} {\bibfnamefont {M.}~\bibnamefont {Polini}},\ }\href
  {\doibase http://dx.doi.org/10.1016/j.ssc.2011.07.015} {\bibfield  {journal}
  {\bibinfo  {journal} {Solid State Comm.}\ }\textbf {\bibinfo {volume}
  {151}},\ \bibinfo {pages} {1627 } (\bibinfo {year} {2011})}\BibitemShut
  {NoStop}%
\bibitem [{\citenamefont {Fei}\ \emph {et~al.}(2015)\citenamefont {Fei},
  \citenamefont {Iwinski}, \citenamefont {Ni}, \citenamefont {Zhang},
  \citenamefont {Bao}, \citenamefont {Rodin}, \citenamefont {Lee},
  \citenamefont {Wagner}, \citenamefont {Liu}, \citenamefont {Dai},
  \citenamefont {Goldflam}, \citenamefont {Thiemens}, \citenamefont {Keilmann},
  \citenamefont {Lau}, \citenamefont {Castro-Neto}, \citenamefont {Fogler},\
  and\ \citenamefont {Basov}}]{Tunneling_plamonics}%
  \BibitemOpen
  \bibfield  {author} {\bibinfo {author} {\bibfnamefont {Z.}~\bibnamefont
  {Fei}}, \bibinfo {author} {\bibfnamefont {E.~G.}\ \bibnamefont {Iwinski}},
  \bibinfo {author} {\bibfnamefont {G.~X.}\ \bibnamefont {Ni}}, \bibinfo
  {author} {\bibfnamefont {L.~M.}\ \bibnamefont {Zhang}}, \bibinfo {author}
  {\bibfnamefont {W.}~\bibnamefont {Bao}}, \bibinfo {author} {\bibfnamefont
  {A.~S.}\ \bibnamefont {Rodin}}, \bibinfo {author} {\bibfnamefont
  {Y.}~\bibnamefont {Lee}}, \bibinfo {author} {\bibfnamefont {M.}~\bibnamefont
  {Wagner}}, \bibinfo {author} {\bibfnamefont {M.~K.}\ \bibnamefont {Liu}},
  \bibinfo {author} {\bibfnamefont {S.}~\bibnamefont {Dai}}, \bibinfo {author}
  {\bibfnamefont {M.~D.}\ \bibnamefont {Goldflam}}, \bibinfo {author}
  {\bibfnamefont {M.}~\bibnamefont {Thiemens}}, \bibinfo {author}
  {\bibfnamefont {F.}~\bibnamefont {Keilmann}}, \bibinfo {author}
  {\bibfnamefont {C.~N.}\ \bibnamefont {Lau}}, \bibinfo {author} {\bibfnamefont
  {A.~H.}\ \bibnamefont {Castro-Neto}}, \bibinfo {author} {\bibfnamefont
  {M.~M.}\ \bibnamefont {Fogler}}, \ and\ \bibinfo {author} {\bibfnamefont
  {D.~N.}\ \bibnamefont {Basov}},\ }\href@noop {} {\bibfield  {journal}
  {\bibinfo  {journal} {Nano Letters}\ }\textbf {\bibinfo {volume} {15}},\
  \bibinfo {pages} {4973} (\bibinfo {year} {2015})}\BibitemShut {NoStop}%
\bibitem [{\citenamefont {Lambe}\ and\ \citenamefont
  {McCarthy}(1976)}]{Lambe_PRL}%
  \BibitemOpen
  \bibfield  {author} {\bibinfo {author} {\bibfnamefont {J.}~\bibnamefont
  {Lambe}}\ and\ \bibinfo {author} {\bibfnamefont {S.~L.}\ \bibnamefont
  {McCarthy}},\ }\href {\doibase 10.1103/PhysRevLett.37.923} {\bibfield
  {journal} {\bibinfo  {journal} {Phys. Rev. Lett.}\ }\textbf {\bibinfo
  {volume} {37}},\ \bibinfo {pages} {923} (\bibinfo {year} {1976})}\BibitemShut
  {NoStop}%
\bibitem [{\citenamefont {Parzefall}\ \emph {et~al.}(2015)\citenamefont
  {Parzefall}, \citenamefont {Bharadwaj}, \citenamefont {Jain}, \citenamefont
  {Taniguchi}, \citenamefont {Watanabe},\ and\ \citenamefont
  {Novotny}}]{Novotny_hBN}%
  \BibitemOpen
  \bibfield  {author} {\bibinfo {author} {\bibfnamefont {M.}~\bibnamefont
  {Parzefall}}, \bibinfo {author} {\bibfnamefont {P.}~\bibnamefont
  {Bharadwaj}}, \bibinfo {author} {\bibfnamefont {A.}~\bibnamefont {Jain}},
  \bibinfo {author} {\bibfnamefont {T.}~\bibnamefont {Taniguchi}}, \bibinfo
  {author} {\bibfnamefont {K.}~\bibnamefont {Watanabe}}, \ and\ \bibinfo
  {author} {\bibfnamefont {L.}~\bibnamefont {Novotny}},\ }\href@noop {}
  {\bibfield  {journal} {\bibinfo  {journal} {Nature Nanotechnology}\ }\textbf
  {\bibinfo {volume} {10}},\ \bibinfo {pages} {1058} (\bibinfo {year}
  {2015})}\BibitemShut {NoStop}%
\bibitem [{\citenamefont {Yadav}\ \emph {et~al.}(2015)\citenamefont {Yadav},
  \citenamefont {Tombet}, \citenamefont {Watanabe}, \citenamefont {Ryzhii},\
  and\ \citenamefont {Otsuji}}]{THz_emission_DGL_experim}%
  \BibitemOpen
  \bibfield  {author} {\bibinfo {author} {\bibfnamefont {D.}~\bibnamefont
  {Yadav}}, \bibinfo {author} {\bibfnamefont {S.}~\bibnamefont {Tombet}},
  \bibinfo {author} {\bibfnamefont {T.}~\bibnamefont {Watanabe}}, \bibinfo
  {author} {\bibfnamefont {V.}~\bibnamefont {Ryzhii}}, \ and\ \bibinfo {author}
  {\bibfnamefont {T.}~\bibnamefont {Otsuji}},\ }in\ \href {\doibase
  10.1109/DRC.2015.7175678} {\emph {\bibinfo {booktitle} {73rd Annual Device
  Research Conference (DRC)}}}\ (\bibinfo {year} {2015})\ pp.\ \bibinfo {pages}
  {271--272}\BibitemShut {NoStop}%
\bibitem [{\citenamefont {Dubinov}\ \emph {et~al.}(2014)\citenamefont
  {Dubinov}, \citenamefont {Aleshkin}, \citenamefont {Ryzhii}, \citenamefont
  {Shur},\ and\ \citenamefont {Otsuji}}]{SP-lasing}%
  \BibitemOpen
  \bibfield  {author} {\bibinfo {author} {\bibfnamefont {A.~A.}\ \bibnamefont
  {Dubinov}}, \bibinfo {author} {\bibfnamefont {V.~Y.}\ \bibnamefont
  {Aleshkin}}, \bibinfo {author} {\bibfnamefont {V.}~\bibnamefont {Ryzhii}},
  \bibinfo {author} {\bibfnamefont {M.~S.}\ \bibnamefont {Shur}}, \ and\
  \bibinfo {author} {\bibfnamefont {T.}~\bibnamefont {Otsuji}},\ }\href@noop {}
  {\bibfield  {journal} {\bibinfo  {journal} {J. Appl. Phys.}\ }\textbf
  {\bibinfo {volume} {115}},\ \bibinfo {eid} {044511} (\bibinfo {year}
  {2014})}\BibitemShut {NoStop}%
\bibitem [{\citenamefont {Falkovsky}\ and\ \citenamefont
  {Varlamov}(2007)}]{Falkovsky-Varlamov}%
  \BibitemOpen
  \bibfield  {author} {\bibinfo {author} {\bibfnamefont {L.~A.}\ \bibnamefont
  {Falkovsky}}\ and\ \bibinfo {author} {\bibfnamefont {A.~A.}\ \bibnamefont
  {Varlamov}},\ }\href {\doibase 10.1140/epjb/e2007-00142-3} {\bibfield
  {journal} {\bibinfo  {journal} {Eur. Phys. J. B}\ }\textbf {\bibinfo {volume}
  {56}},\ \bibinfo {pages} {281} (\bibinfo {year} {2007})}\BibitemShut
  {NoStop}%
\bibitem [{\citenamefont {Bhatnagar}\ \emph {et~al.}(1954)\citenamefont
  {Bhatnagar}, \citenamefont {Gross},\ and\ \citenamefont
  {Krook}}]{BGK-collisions}%
  \BibitemOpen
  \bibfield  {author} {\bibinfo {author} {\bibfnamefont {P.~L.}\ \bibnamefont
  {Bhatnagar}}, \bibinfo {author} {\bibfnamefont {E.~P.}\ \bibnamefont
  {Gross}}, \ and\ \bibinfo {author} {\bibfnamefont {M.}~\bibnamefont
  {Krook}},\ }\href {\doibase 10.1103/PhysRev.94.511} {\bibfield  {journal}
  {\bibinfo  {journal} {Phys. Rev.}\ }\textbf {\bibinfo {volume} {94}},\
  \bibinfo {pages} {511} (\bibinfo {year} {1954})}\BibitemShut {NoStop}%
\end{thebibliography}%

\end{document}